\documentclass[sigconf,nonacm]{acmart}
\usepackage{amsmath,amssymb,amsfonts}
\usepackage{algorithmic}
\usepackage{algorithm}
\usepackage{graphicx}
\usepackage{textcomp}
\usepackage{xcolor}
\usepackage{subfigure}
\usepackage{bm}
\usepackage{threeparttable}
\usepackage{tabularx}

\AtBeginDocument{%
  \providecommand\BibTeX{{%
    \normalfont B\kern-0.5em{\scshape i\kern-0.25em b}\kern-0.8em\TeX}}}
    
\setcopyright{acmcopyright}
\copyrightyear{2020}
\acmYear{2020}
\acmDOI{10.1145/1122445.1122456}

\acmConference[Mobihoc '20]{Mobihoc '20: ACM International Symposium on Theory, Algorithmic Foundations, and Protocol Design for Mobile Networks and Mobile Computing}{June 30-- July 03, 2020}{Shanghai, China}
\acmBooktitle{Woodstock '20: ACM International Symposium on Theory, Algorithmic Foundations, and Protocol Design for Mobile Networks and Mobile Computing,
  June 30-- July 03, 2020, Shanghai, China}
\acmPrice{15.00}

\begin{document}

\title{MVStylizer: An Efficient Edge-Assisted Video Photorealistic Style Transfer System for Mobile Phones}

\author{Ang Li}
\affiliation{\institution{Duke University}}
\email{ang.li630@duke.edu}

\author{Chunpeng Wu}
\affiliation{\institution{Duke University}}
\email{chunpeng.wu@duke.edu}

\author{Yiran Chen}
\affiliation{\institution{Duke University}}
\email{yiran.chen@duke.edu}

\author{Bin Ni}
\affiliation{\institution{Quantil Inc.}}
\email{nibin@quantil.com}

\renewcommand{\shortauthors}{A. Li, et al.}

\begin{abstract}
Recent research has made great progress in realizing neural style transfer of images, which denotes transforming an image to a desired style. 
Many users start to use their mobile phones to record their daily life, and then edit and share the captured images and videos with other users. 
However, directly applying existing style transfer approaches on videos, i.e., transferring the style of a video frame by frame, requires an extremely large amount of computation resources. 
It is still technically unaffordable to perform style transfer of videos on mobile phones.
To address this challenge, we propose MVStylizer, an efficient edge-assisted photorealistic video style transfer system for mobile phones. 
Instead of performing stylization frame by frame, only key frames in the original video are processed by a pre-trained deep neural network (DNN) on edge servers, while the rest of stylized intermediate frames are generated by our designed optical-flow-based frame interpolation algorithm on mobile phones. 
A meta-smoothing module is also proposed to simultaneously upscale a stylized frame to  arbitrary  resolution and remove style transfer related distortions in these upscaled frames. In addition, for the sake of continuously enhancing the performance of the DNN model on the edge server, we adopt a federated learning scheme to keep retraining each DNN model on the edge server with collected data from mobile clients and syncing with a global DNN model on the cloud server.
Such a scheme effectively leverages the diversity of collected data from various mobile clients and efficiently improves the system performance. 
Our experiments demonstrate that MVStylizer can generate stylized videos with an even better visual quality compared to the state-of-the-art method while achieving 75.5$\times$ speedup  for 1920$\times$1080 videos.

\end{abstract}



\maketitle

\section{Introduction}
\begin{figure}[ht]
    \centering
    \includegraphics[scale=0.28]{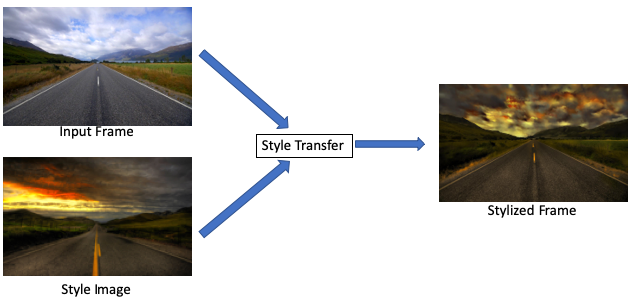}
    \caption{An example of video style transfer.}
    \label{fig:example}
\end{figure}

In the past decade, deep neural networks (DNNs) have been widely applied in image transformation tasks, including style transfer \cite{ulyanov2016texture,johnson2016perceptual,gatys2015neural}, semantic segmentation \cite{long2015fully}, super resolution \cite{dong2015image,johnson2016perceptual}, etc. DNN-based style transfer is one of the most popular techniques in image transformation, and has led to many successful industrial applications with significant commercial impacts, such as Prisma \cite{prisma} and DeepArt \cite{deepart}. The DNN-based style transfer aims at transforming an input image into a desired output image according to a user-specified style image. Specifically, the DNN model is trained to search for a new image that has similar neural activations as the input image's and similar feature correlations as the style image's. Figure \ref{fig:example} shows one example of directly applying a pre-trained DNN model to perform style transfer. Here the input image is one extracted frame from a video of road trip recorded in the daytime, while the style image is a similar scene captured at dusk. After performing stylization, the input frame is successfully transformed to the dusky scene while keeping the content unchanged as the input frame.

\begin{figure*}[t]
\centering
\makebox[20pt]{\raisebox{30pt}{\rotatebox[origin=c]{90}{input image}}}
\subfigure{\includegraphics[height=2.2cm,width=3cm]{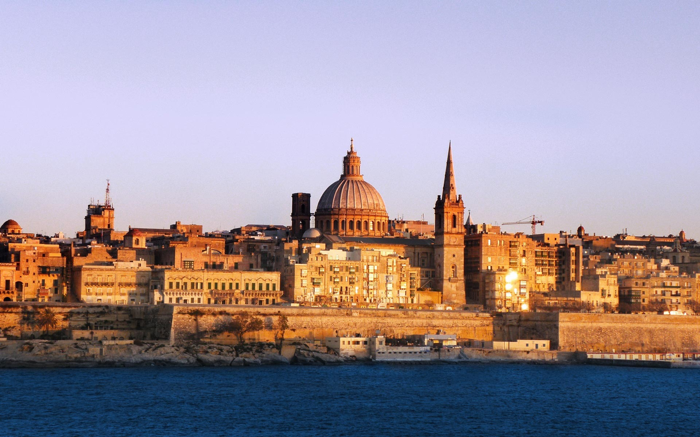}}
\subfigure{\includegraphics[height=2.2cm,width=3cm]{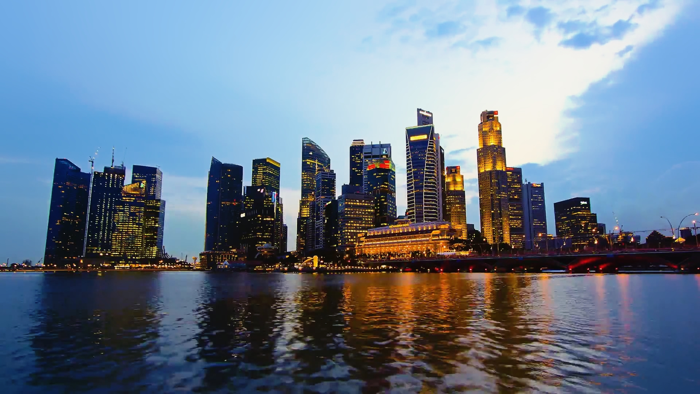}}
\subfigure{\includegraphics[height=2.2cm,width=3cm]{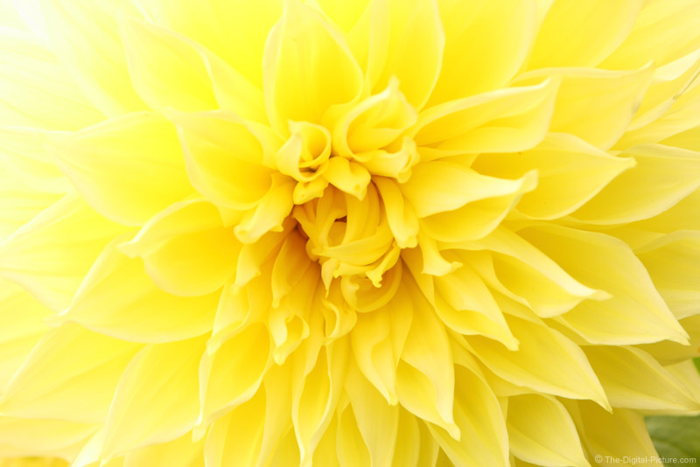}}
\subfigure{\includegraphics[height=2.2cm,width=3cm]{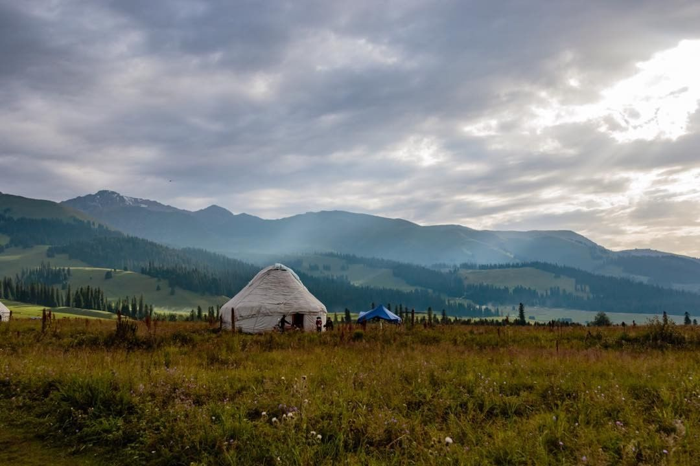}}
\subfigure{\includegraphics[height=2.2cm,width=3cm]{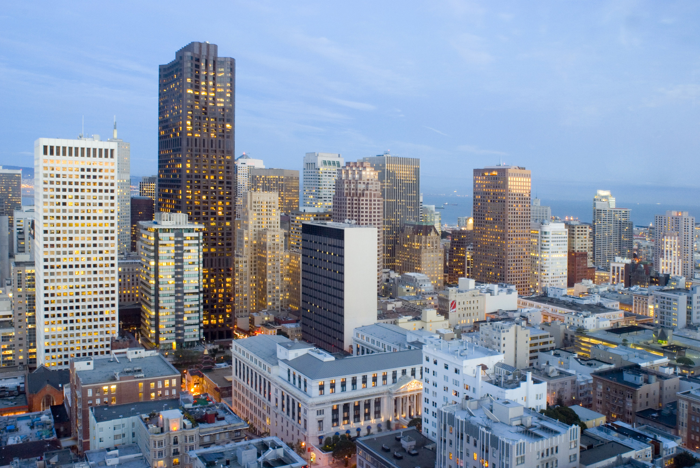}}
\\
\makebox[20pt]{\raisebox{30pt}{\rotatebox[origin=c]{90}{style image}}}
\subfigure{\includegraphics[height=2.2cm,width=3cm]{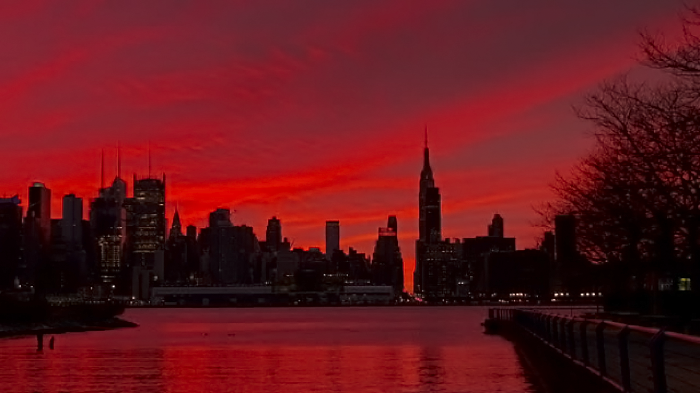}}
\subfigure{\includegraphics[height=2.2cm,width=3cm]{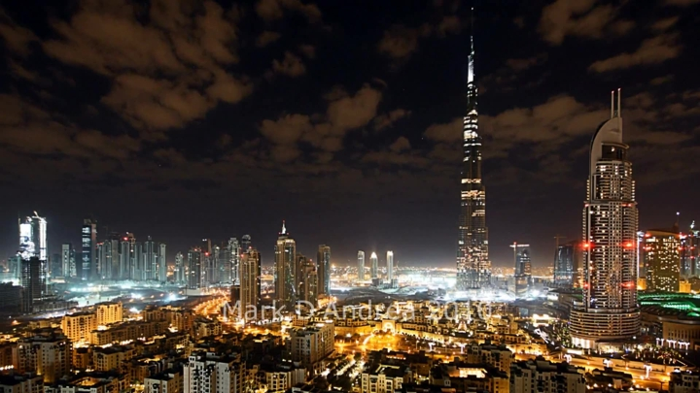}}
\subfigure{\includegraphics[height=2.2cm,width=3cm]{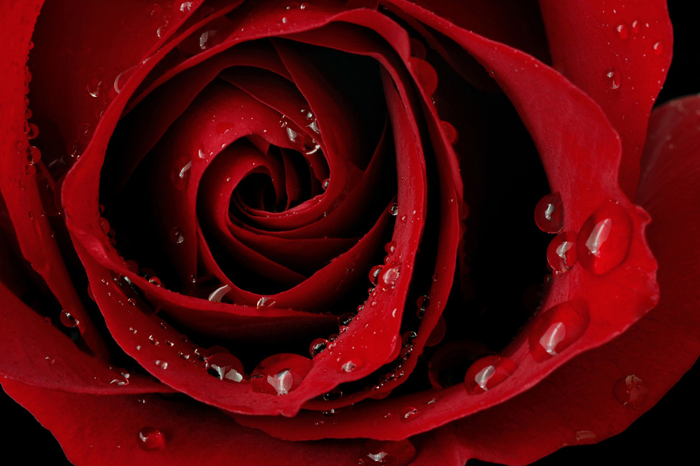}}
\subfigure{\includegraphics[height=2.2cm,width=3cm]{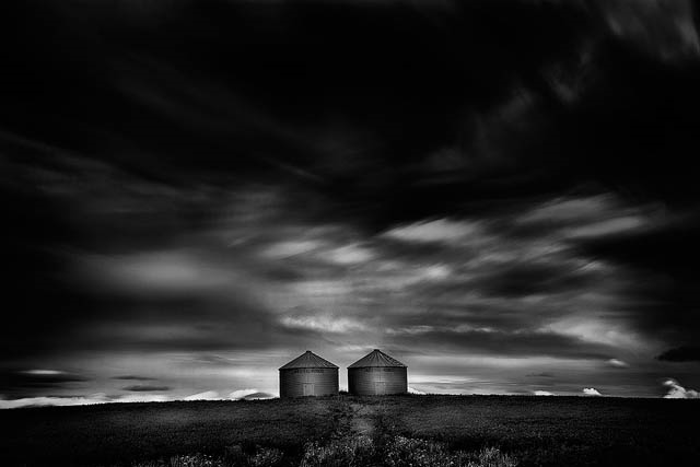}}
\subfigure{\includegraphics[height=2.2cm,width=3cm]{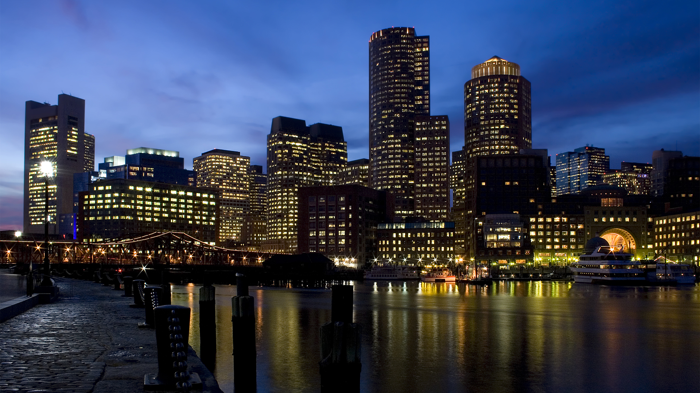}}\\
\makebox[20pt]{\raisebox{30pt}{\rotatebox[origin=c]{90}{artistic}}}
\subfigure{\includegraphics[height=2.2cm,width=3cm]{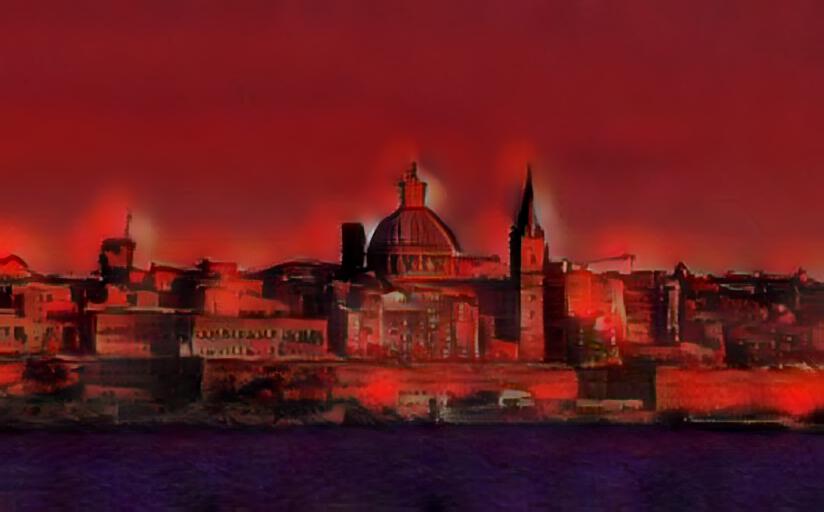}}
\subfigure{\includegraphics[height=2.2cm,width=3cm]{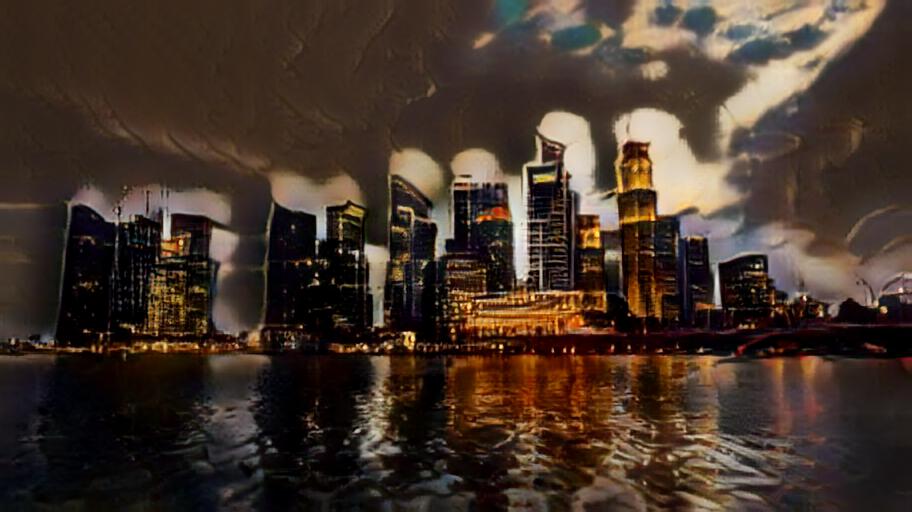}}
\subfigure{\includegraphics[height=2.2cm,width=3cm]{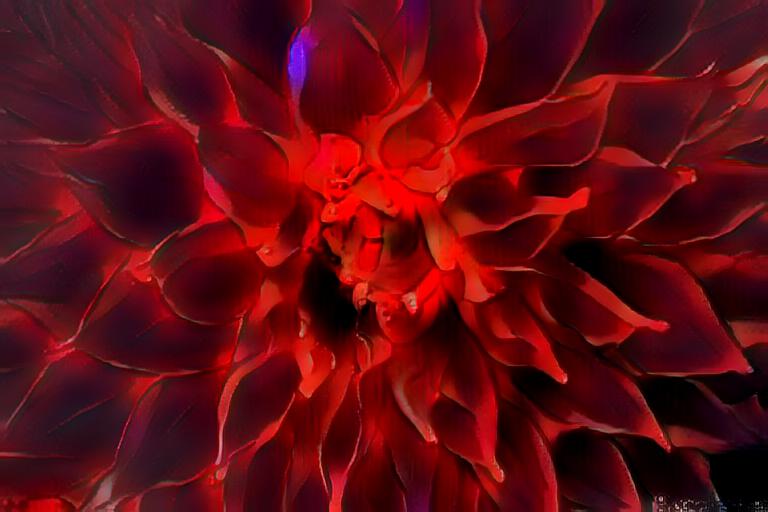}}
\subfigure{\includegraphics[height=2.2cm,width=3cm]{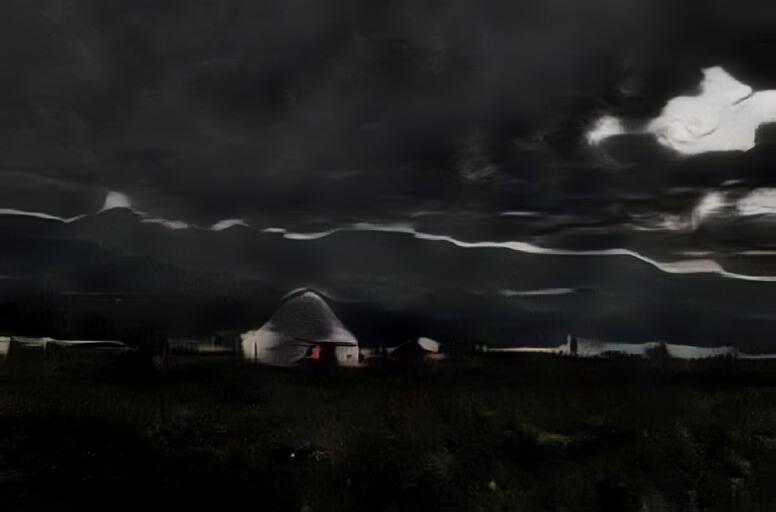}}
\subfigure{\includegraphics[height=2.2cm,width=3cm]{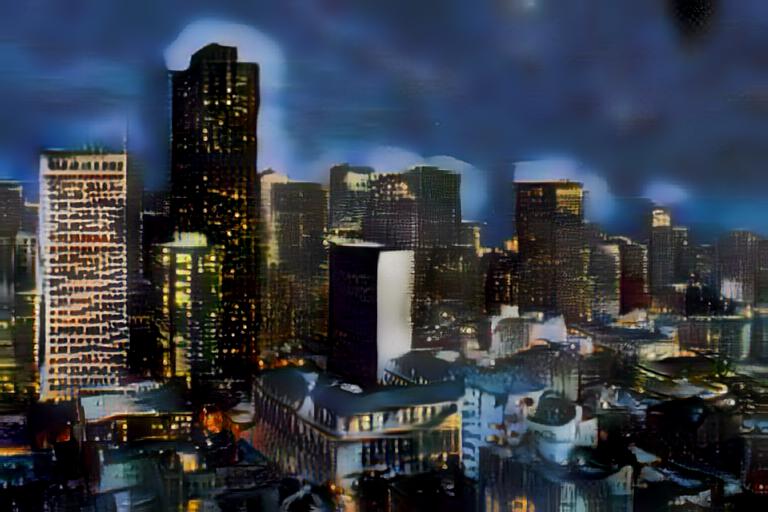}}\\
\makebox[20pt]{\raisebox{30pt}{\rotatebox[origin=c]{90}{photorealistic}}}
\subfigure{\includegraphics[height=2.2cm,width=3cm]{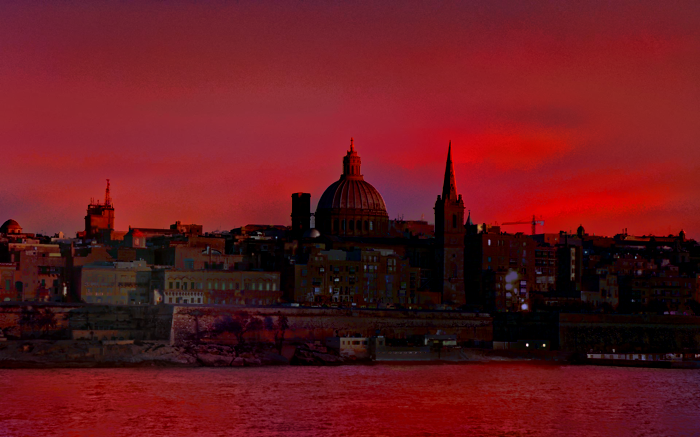}}
\subfigure{\includegraphics[height=2.2cm,width=3cm]{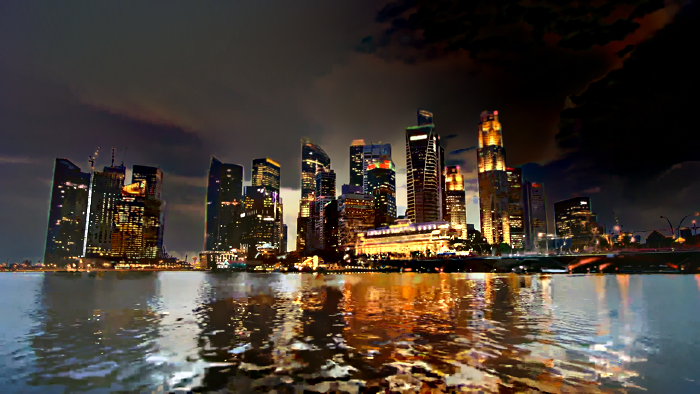}}
\subfigure{\includegraphics[height=2.2cm,width=3cm]{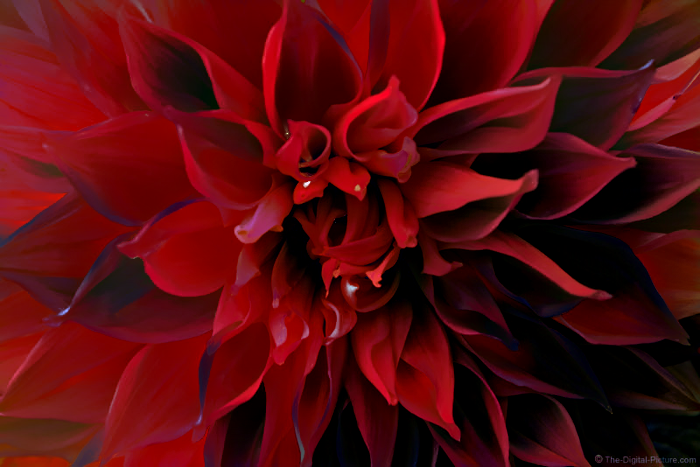}}
\subfigure{\includegraphics[height=2.2cm,width=3cm]{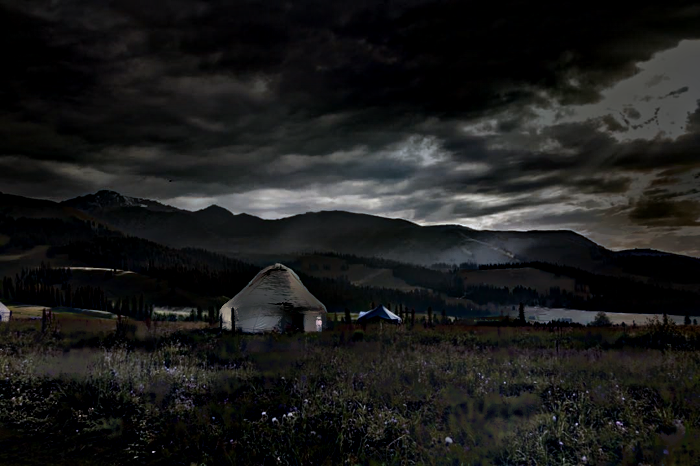}}
\subfigure{\includegraphics[height=2.2cm,width=3cm]{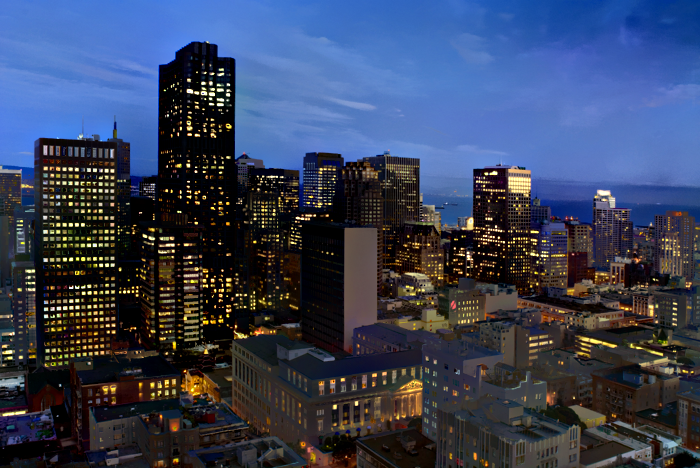}}\\

\caption{Examples of artistic style transfer \cite{huang2017adain}  and photorealistic style transfer \cite{luan2017deep}.}\label{fig:example1}
\end{figure*}

However, the naive extension of style transfer from images to videos is very challenging: frame-by-frame transformations are very slow \cite{chen2017coherent,huang2017real} even running on powerful GPUs. 
Although many users start to use mobile phones to record their daily life, and then edit and share images and videos on social networks or with friends, performing style transfer of videos on mobile phones is still unaffordable till very recently due to the limited computing sources on the phones. 
In addition to efficiency issues, ``photorealistic'' style transfer is another critical challenge. Figure \ref{fig:example1} shows several examples of artistic style transfer and photorealistic style transfer. Even though the contents of artistic stylized images are distorted, these distortions can be tolerated and hard to be detected by human eyes due to the artistic attribute. However, compared with artistic style transfer, the target of photorealistic style transfer is to achieve photorealism, which requires loyally preserving the content structure in the stylized image. Humans are able to evaluate the visual quality of the photorealistic stylized images. 
It is necessary to explore the approach that can achieve efficiently photorealistic style transfer of videos on mobile phones while keeping high visual quality of the stylized videos.
Some works have been done for style transfer of images and videos, most of which focus on improving either the visual quality of stylization \cite{anderson2016deepmovie,ruder2016artistic,lee2010directional,li2016combining} or efficiency of style transfer  \cite{chen2017stylebank,dumoulin2016learned,huang2017real,li2017diversified,ulyanov2016texture}. However, performing photorealistic style transfer of videos on mobile phones under the constraints of computing resources has not been fully investigated.

In this paper, we propose MVStylizer -- an  efficient edge-assisted video style transfer system  for mobile phones. Instead of performing transformation frame by frame, MVStylizer processes only extracted key frames using pre-trained DNN models on the edge server while the rest of intermediate frames are generated on-the-fly using our proposed optical-flow-based frame interpolation algorithm on mobile phones. 
The interpolation is done by exploiting the optical flow information between the intermediate frames and key frames. The reason why we choose edge servers but not cloud servers is that edge servers are closer to users and hence, are able to provide real-time response to mobile phones. 
Edge-cloud federated learning is adopted in our architecture to continuously improve the performance of the DNN models on edge server:
Each DNN model on the edge server keeps re-training while performing stylization with collected data from mobile clients, and each edge server will sync the model with cloud server where a global DNN model is maintained when it is idle. Meanwhile, the global DNN is also a backup of the edge DNN, from which an edge server can be quickly restored if it suddenly crashes. We also conduct  experiments to quantitatively and qualitatively evaluate our proposed system. The experimental results demonstrate that MVStylizer can successfully stylize videos with even better visual quality compared to the state-of-the-art method while achieving significant speedup with high resolutions.

The main contributions of this paper are summarized as follows:
\begin{itemize}
    \item To the best of our knowledge, MVStylizer is the first mobile system for performing photorealistic  style transfer of videos.
    \item An optical-flow-based frame interpolation algorithm is proposed to accelerate style transfer of videos on mobile phones.
    \item A meta-smoothing module is  designed to efficiently tackle two problems in an end-to-end learning manner: dynamically upscaling a stylized image to multiple/arbitrary resolution and removing style transfer related distortions in these upscaled versions.
    \item An edge-cloud federated learning scheme is applied to continuously improve the performance of DNN-based stylizer on edge servers. 
    \item We implement a prototype system and conduct  experiments to quantitatively evaluate  MVStylizer, which demonstrates that it can perform video style transfer of videos on mobile phones in an effective and efficient way.
\end{itemize}

The rest of this paper is organized as follows. Section \ref{sec:backgroud} provides background information and section \ref{sec:design} presents the system overview and details of core modules.  Section \ref{sec:experiment} shows the evaluation results. Section \ref{sec:related} reviews the related work. Section \ref{sec:conclusion} concludes this paper.

\section{Background} \label{sec:backgroud}

\subsection{Photorealistic style transfer}
As Figure \ref{fig:example} shows, the photorealistic style transfer extracts human-perceptual ``style'' features from  a style image and applies these style features to ``decorate'' the content image without changing objects' semantic structures in the original photo.
Similar to artistic style transfer~\cite{2017arXiv170504058J}, photorealistic style transfer requires high style faithfulness for the stylized image. 
Unlike artistic style transfer, however, generating a stylized image with high photorealism is essential in photorealistic style transfer, which remains as a key challenge. 
%
Based on previous artistic style transfer methods~\cite{johnson2016perceptual,Li_NeurIPS17}, theoretical analysis on feature correlation~\cite{2018arXiv180206474L} or photorealism loss~\cite{luan2017deep} are often adopted to quantitatively approximate or simulate human evaluation on photorealism.

\subsection{Device to edge offloading}
Offloading computational intensive tasks to edge servers is feasible and promising way to address the challenge of limited computation resources on edge devices (e.g., smartphones) \cite{xu2020survey}. Such strategy has been widely applied to many applications. Ran \textit{et al.} \cite{ran2018deepdecision} design a framework to dynamically determine the offloading strategy for the object detection task based on the network conditions.  Yi \textit{et al.} \cite{yi2017lavea} propose a system named Lavea, which offloads the computation from the clients to nearby edge servers to provide video analytics service with low latency. Jeong \textit{et al.} \cite{jeong2018computation} propose an approach to offload DNN computations to nearby edge servers in the context of web apps. Chen \textit{et al.} \cite{chen2017empirical} conduct an empirical study to evaluate the performance of several edge computing applications in terms of latency. 

In this work, we also adopt the similar strategy to offload key frames to the edge servers, where those key frames are processed by the pre-trained DNN models. We leave the lightweight optical-flow-based interpolation for intermediate frames on mobile phones.

\section{Our Proposed MVStylizer System}\label{sec:design}

\subsection{The System Overview}
In this work, we propose MVStylizer for efficiently performing photorealistic style transfer for videos on mobile phones. Due to the constrained computation resources on mobiles, edge servers are leveraged to speed up the style transfer. 
Moreover, specific technical approaches are proposed to address two critical challenges in this system.

First, performing frame-by-frame stylization is still technically unaffordable, even with the assistance of edge servers. 
We propose an optical-flow-based frame interpolation algorithm and a meta-smoothing module to speed up the stylization process. 
Specifically, only extracted key frames will be processed by a pre-trained DNN on the edge server, while the rest of intermediate frames will be generated on-the-fly using our proposed optical-flow-based frame interpolation algorithm on mobile phones. 
The interpolation is done based on the stylized key frames and pre-computed optical flow information between key frames and intermediate frames. 
In addition, the meta-smoothing module is integrated in the edge DNN for handling upscaling and distortion issues of the stlyized frame in a single operation, accelerating the style transfer of key frames on the server. 

Second, the edge DNN may be trained with limited data so that the optimal performance is not achieved. 
Therefore, we propose an edge-cloud federated  learning scheme to continuously improve the performance of the edge DNN. 
While the edge server offers stylization service, the edge DNN keeps retraining based on the collected data from mobile clients. 
The updated edge DNN will be synced with a cloud DNN when the edge server is idle. 
Note that the cloud DNN  has the same DNN architecture as edge DNNs, and the cloud DNN is also maintained as a backup of the edge DNN in case some edge server crashes. 
The cloud server is updated with averaging the aggregated parameters from each server, and the updated parameters of the cloud DNN will be synced with each edge DNN.

\begin{figure}[t]
\centering
\includegraphics[width=\columnwidth]{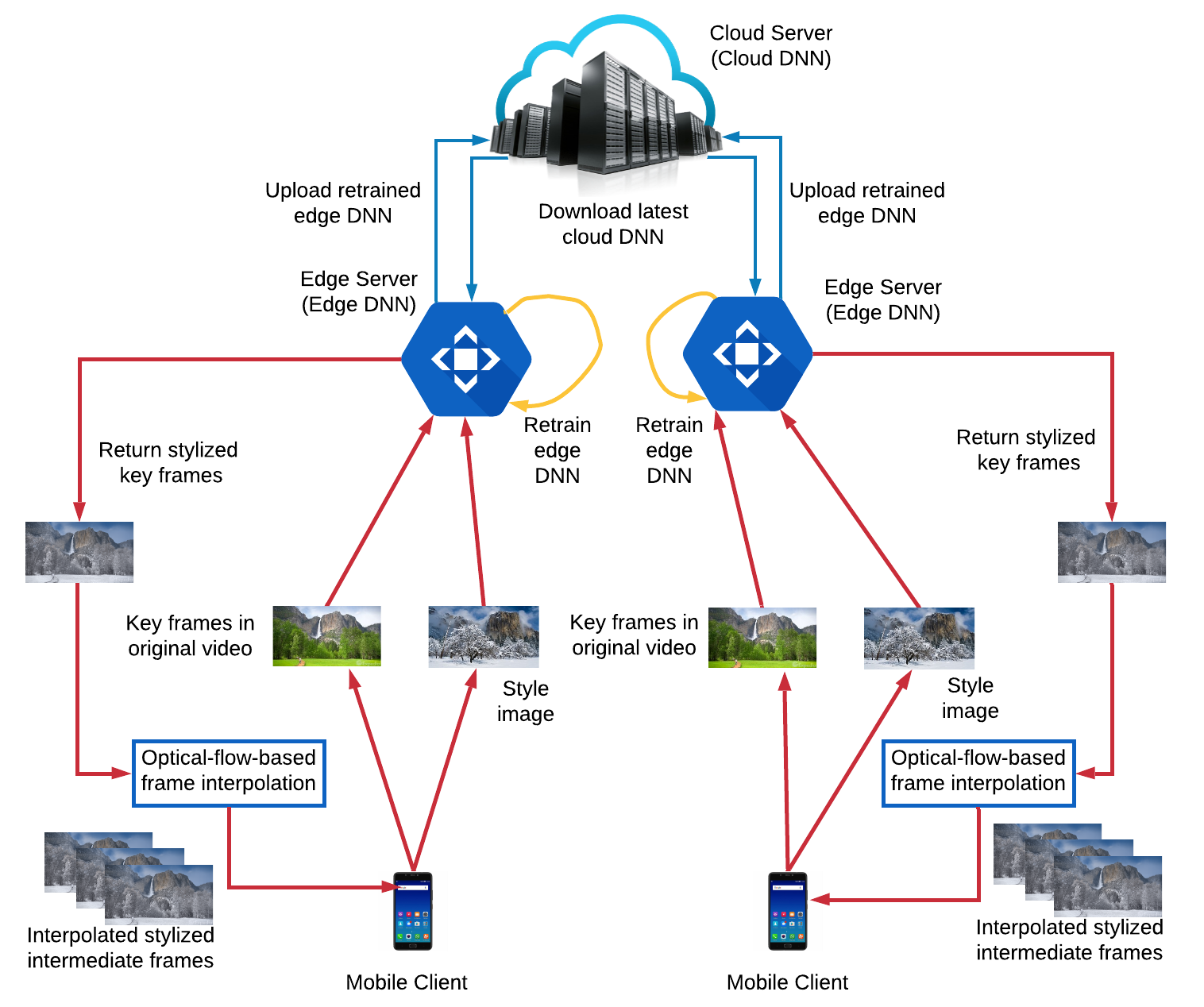}
\caption{The system design of MVStylizer.}
\label{fig:design}
\end{figure}

As illustrated in Figure \ref{fig:design}, MVStylizer consists of three major modules: \textit{optical-flow-based frame interpolation}, \textit{edge DNN} and \textit{cloud DNN}. 
The work flow is as follows: 
When a user performs the style transfer of a video on the mobile phone, the extracted key frames of this video will be sent to the edge server associated with a user-specified style image. The key frames can be identified using either standard H.264 video codec or the content-based method \cite{Mao:2019:MEC:3316781.3317865}.
Afterwards, a pre-trained DNN will perform transformation on received key frames according to the style image on the edge server. 
While performing stylization, the edge DNN will continuously keep re-training based on the specific evaluation metric. 
The updated edge DNN will be synced with a cloud DNN when the edge server is idle.
The post-processed stylized key frames will then be returned to the mobile client. 
Finally, the stylized intermediate frames will be interpolated by the proposed optical-flow-based frame interpolation algorithm.

Next, we will illustrate more details about each module.

\subsection{Optical-flow-based Frame Interpolation}

Transferring the style of a video on a mobile phone is always challenging due to its limited available computing resource.
In this work, we attempt to process only a few key frames on the edge DNN, while a lot of more stylized intermediate frames will be interpolated based on the optical flow information. 
An optical-flow-based frame interpolation algorithm is designed for interpolating stylized intermediate frames based on the optical flow information between intermediate frames and key frames in the original video.

Given a video consists of $n$ frames, among which there are $j$ key frames $(k_0,\ldots,k_{j-1})$ and $m$ intermediate frames $(i_0,\ldots,i_{m-1})$ where $j+m=n$. 
The optical flow information can be computed as $f_{sp}=F(i_s,k_p)$ for any intermediate frame $i_s$ between the key frame $k_p$ and $k_{p+1}$, where $F$ is an optical flow estimator \cite{lucas1981iterative}. 
The $f_{sp}$ is a two dimensional flow field, representing the displacement of each pixel from $k_p$ to $i_s$. 
For example, given the location of a pixel \textbf{p} in $k_p$ as $k_p(x,y)$, and assume the location of \textbf{p} in $i_s$ is $k_p(x+\Delta x,y+\Delta y)$, that is, $i_s(x,y)=k_p(x+\Delta x,y+\Delta y)$. 
The optical flow of \textbf{p} from $k_p$ to $i_s$ is $(\Delta x, \Delta y)$. 
Again, only key frames will be sent to the edge server for performing style transfer using the edge DNN, and the stylized key frames $(\hat{k}_0,\ldots,\hat{k}_{j-1})$ will be returned to the mobile phone. 
Based on the stylized key frames $(\hat{k}_0,\ldots,\hat{k}_{j-1})$ and optical flow information $(f_0,\ldots,f_{m-1})$, we are able to generate stylized intermediate frames $(\hat{i}_0,\ldots,\hat{i}_{m-1})$ by spatial warping. 
We adopt the bilinear interpolation for generating the stylized intermediate frames, such as:
\begin{equation}
\hat{i}_p=I(f_{pq},\hat{k}_q),
\label{eq:bilinear}
\end{equation}
where $I$ is a bilinear interpolation kernel. The detailed workflow of the optical-flow-based interpolation algorithm is described in \textbf{Algorithm \ref{alg:Framwork}}.

\begin{algorithm}[ht]   
	\caption{Optical-flow-based frame interpolation}   
	\label{alg:Framwork}   
	\begin{algorithmic}[1] 
		\REQUIRE ~~\\ 
		Stylized key frames $(\hat{k}_0,\ldots,\hat{k}_{j-1})$;\\
		Optical flow information between key frames and intermediate frames $(f_0,\ldots,f_{m-1})$;\\
		Index of each frame in original video 
		\ENSURE ~~\\ Stylized intermediate frames $(\hat{i}_0,\ldots,\hat{i}_{m-1})$; \\
		\STATE $p=0$, $q=0$
		\STATE \textbf{for} p = 0 : m:
		\STATE \hspace{0.5cm} \textbf{if} 
		Index ($\hat{k}_q$) $<$ Index ($f_p$) $<$ Index ($\hat{k}_{q+1}$)
		\STATE \hspace{0.8cm} $\hat{i}_p=I(f_p,\hat{k}_q)$;
		\STATE \hspace{0.5cm} \textbf{else} 
		\STATE \hspace{0.8cm} $q = q+1$
		\STATE \hspace{0.5cm} \textbf{end if}
		\STATE \textbf{end for}
	\end{algorithmic}
\end{algorithm}

\subsection{Edge DNN}

\begin{figure}[ht]
\centering
\includegraphics[width=\columnwidth]{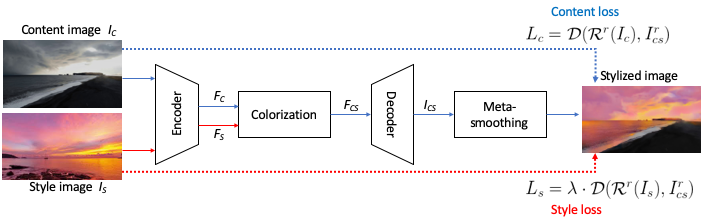}
\caption{The design of DNN-based stylizer with meta-smoothing.}
\label{fig:stylizer}
\end{figure}

In MVStylizer, the edge DNN is a stylizer. 
As depicted in Figure~\ref{fig:stylizer}, it consists of four modules: a pretrained VGG~\cite{Simonyan_ICLR15} encoder for feature extraction, a colorization module for integrating style features into content features, a decoder with mirrored VGG layers for generating stylized image, and our proposed \textit{meta-smoothing module}.

The meta-smoothing module is designed to tackle two major issues in an \textit{end-to-end} learning manner: dynamically upscaling the decoder's output to multiple/arbitrary resolution and removing the style transfer related distortions in these upscaled versions.
First, popular encoder-decoder for style transfer~\cite{Huang_ICCV17,Li_NeurIPS17} often adopt fixed architectures (VGG, ResNet~\cite{He_CVPR16}, MobileNet~\cite{Howard_arxiv17}, \textit{etc}.).
Even when replacing the encoder-decoder with a more flexible DNN transformer~\cite{Shen_CVPR18} to support manually designed architecture, 
the generated image still needs to be upscaled to the target resolution, and the upscaling efficiency is an issue.
Second, the stylized image generated by the decoder has distortions incurred by the style transfer. 
Those distortions become even worsened in the upscaling process as observed in previous super resolution studies~\cite{Hu_CVPR19,Shi_CVPR16}.
In other words, adopting only super resolution methods cannot synthesize a satisfactory stylized image with high resolution.

In our method, as shown in Figure \ref{fig:stylizer}, a content image $\bm{I}_{c}$ and a style image $\bm{I}_{s}$ are rescaled to the same size and fed into our stylizer's pretrained VGG encoder. 
The encoder will extract the content features $\bm{F}_{c}$ (blue) and style features $\bm{F}_{s}$ (red). 
After that, the colorization module will ``colorize'' $\bm{F}_{c}$ with $\bm{F}_{s}$. 
The feature maps $\bm{F}_{c}$ and $\bm{F}_{s}$ have the same dimension, since the input $\bm{I}_{c}$ and $\bm{I}_{s}$ have been rescaled to the same size.
The dimensions of $\bm{F}_{c}$ and $\bm{F}_{s}$ are denoted as $H \times W \times C$ where $H \times W$ is the feature map's height and width and $C$ is the channel number.
The colorization is processed along channel dimensions of $\bm{F}_{c}$ and $\bm{F}_{s}$.
Specifically, the feature maps $\bm{F}_{c}$ is denoted as $\{\bm{x}^{i}_{c}|i=1,2,...,H\times W\}$ where the vector length $|\bm{x}^{i}|=C$.
Similarly, the feature maps $\bm{F}_{s}$ is denoted as $\{\bm{y}^{j}_{s}|j=1,2,...,H\times W\}$ where the vector length $|\bm{y}^{j}|=C$.
The colorization module outputs the colorized content features $\bm{F}{cs}$, which is denoted as $\{\bm{z}^{k}_{cs}|z=1,2,...,H\times W\}$ and the vector length $|\bm{z}^{k}|=C$.
The colorization is defined as:
\begin{equation}
\label{eq:our_colorization}
\bm{z}^{k}_{cs}=\sum^{H \times W}_{m=1}(\bm{y}^{m}_{s}-\bm{x}^{k}_{c}).
\end{equation}
Finally, $\bm{F}{cs}$ is fed into the decoder, which generates a stylized image denoted as $\bm{I}_{cs}$. 
The meta-smoothing module upscales and smoothes $\bm{I}_{cs}$ such as
\begin{equation}
\label{eq:our_meta_smoothing}
\bm{I}^{r}_{cs}=\mathcal{P}(\bm{I}_{CS}, \bm{W}^{r}_{u})*\bm{W}_{s},
\end{equation}
where $\mathcal{P}(\cdot)$ is a deconvolution operator, $r$ ($r>0$) is an upscaling factor that is specified by an application user, $\bm{W}^{r}_{u}$ is our improved convolution kernel which is used in the deconvolution operator $\mathcal{P}(\cdot)$, $\bm{W}_{s}$ is convolution kernel $\bm{W}_{s}$ for smoothing distortions incurred by the style transfer, and $N_{s}$ is number of feature maps output by convolution with $\bm{W}_{s}$.
The convolution kernel $\bm{W}^{r}_{u}$ for deconvolution is defined in Equation~\ref{eq:our_conv_kernel}:
\begin{equation}
\label{eq:our_conv_kernel}
\bm{W}^{r}_{u}=
\begin{cases}
\bm{W}^{1}_{u} & r=1\\
\mathcal{Q}(r)*\bm{W}^{1}_{u} & r\neq 1
\end{cases}
\end{equation}
where the function $\mathcal{Q}(\cdot)$ is to construct a matrix filled with $r$. If the upscaling factor $r=1$, the deconvolution operator $\mathcal{P}(\cdot)$ only executes convolution without upscaling.
Therefore, $\bm{W}^{1}_{u}$ is denoted as a meta convolution kernel of our meta-smoothing module.
If the upscaling factor $r\neq 1$, $\bm{W}^{r}_{u}$ is calculated using the meta convolution kernel $\bm{W}^{1}_{u}$ and the upscaling factor $r$.
For $\bm{W}^{r}_{u}$ and $\bm{W}^{s}$, each filter's dimensions are respectively denoted as $H_{u}\times W_{u}$ and $H_{s} \times W_{s}$.

The objective function is defined as Equation~\ref{eq:our_loss}:
\begin{equation}
\label{eq:our_loss}
L=\underbrace{\mathcal{D}(\mathcal{R}^{r}(\bm{I}_{c}), \bm{I}^{r}_{cs})}_{\text{content loss}}+\lambda \cdot \underbrace{\mathcal{D}(\mathcal{R}^{r}(\bm{I}_{s}), \bm{I}^{r}_{cs})}_{\text{style loss}},
\end{equation}
where the function $\mathcal{D}(\cdot)$ is to measure the perceptual distance~\cite{johnson2016perceptual}, the function $\mathcal{R}^{r}(\cdot)$ is to rescale input data with respect to the upscaling factor $r$, and $\lambda$ is a scale factor.
$\mathcal{R}^{r}(\bm{I}_{c})$, $\mathcal{R}^{r}(\bm{I}_{s})$, and $\bm{I}^{r}_{cs}$ have the same size.
In Equation~(\ref{eq:our_loss}), the first part evaluates the \textit{content loss} defined as the perceptual distance between the input content image and the stylized image, and the second part evaluates the \textit{style loss}, which is the perceptual distance between the input style image and the stylized image.

\subsection{Cloud DNN}
The cloud DNN has an identical architecture as the edge DNN. 
It is designed for aggregating the updated parameters of edge DNNs in the edge-cloud federated learning process. 
Algorithm \ref{alg:federated} presents the details of the edge-cloud federated learning procedure. 
Suppose that there exist $N$ participated edge servers, the parameters of each corresponding edge DNN are denoted as ($\theta_1,\cdots,\theta_N$).
$\bar{\theta}$ represents the parameters of the cloud DNN. 
When a mobile client sends a style transfer request to an edge server associated with the video data, the edge DNN will perform the style transfer on received data while retraining the model based on those data. 
Then, the updated parameters $\theta_i$ will be uploaded to the cloud server when the edge server is idle. 
The parameters of the cloud DNN can be updated by averaging the parameters of each edge DNN as $\bar{\theta}^t = \sum_{i=1}^{N} \theta_i^{t}$. 
Finally, the updated $\bar{\theta}$ will be distributed to each edge server, where the edge DNN can be updated with the latest $\bar{\theta}$. 
The entire process will be continuously repeated for improving the performance of the edge DNN in an efficient way. The effectiveness of the edge-cloud federated learning is evaluated in Section \ref{subsec:eval_federated}.

\begin{algorithm}[]   
	\caption{ Edge-Cloud Federated Learning}   
	\label{alg:federated}   
	\begin{algorithmic}[1] 
		\REQUIRE ~~\\ 
		$N$ collected video data  $\mathcal{D}_1, \mathcal{D}_2, \mathcal{D}_3, \dots, \mathcal{D}_N$ on each edge server;
		\ENSURE ~~\\ federated learned neural network parameters $ \bar{\theta} $; 
		\STATE Deploy the pre-trained DNN model on the cloud server;
		\STATE Distribute the cloud DNN to each server as edge DNN for initialization;
		\STATE Initialize time stamp $t=0$;
		\STATE \textbf{Client executes}:
		\STATE \hspace{0.5cm}\textbf{for} $i = 1 : N$:
		\STATE \hspace{1cm}Participated edge server receives latest model parameters $\bar{\theta}^t$ from the cloud server;
		\STATE \hspace{1cm}Update corresponding edge DNN $\theta_{i}^{t+1}=\bar{\theta}^t$;
		\STATE \hspace{1cm}$t = t + 1$;
		\STATE \hspace{1cm}Retrain the edge DNN based on $\mathcal{D}_i$ and send the updated parameters $\theta_{i}^t$;
		\STATE \textbf{Server executes}:
		\STATE \hspace{0.5cm}Compute average parameters $\bar{\theta}^t = \sum_{i=1}^{N} \theta_i^{t}$;
		\STATE \hspace{0.5cm}Send aggregated parameters $\bar{\theta}^t$ to each participated edge server;
		\STATE Continuously repeat from steps 4 to 12 for improving the performance of edge DNN;
	\end{algorithmic}
\end{algorithm}

Not only works for the federated learning, the cloud DNN is also maintained as a backup of the edge DNN, from which an edge server can be quickly restored after it suddenly crashes.
\section{Evaluations}\label{sec:experiment}

\subsection{Experiment Setup}
We implement a prototype system of MVStylizer which is composed of Google Pixel 2 and servers running on Ubuntu 16.04. The mobile device is equipped with Qualcomm Snapdragon-835@2.35GHz. In this experiment, both the edge server and cloud server are  equipped with an Intel Xeon E5-2630@2.6GHz, 128G RAM and a NVIDIA TITAN X Pascal GPU. But the cloud server should be more powerful than the edge server in the real life. The mobile device communicates with the server through IEEE 802.11 wireless network. The system is implemented with PyTorch and OpenCV.

\subsection{Model Training}
To train our proposed stylizer shown in Figure~\ref{fig:stylizer}, the pretrained VGG encoder is frozen, while the other three modules are learned together.
The stylizer is optimized using Adam~\cite{Kingma_ICLR15} with  parameters: $\beta_{1}=0.5$, $\beta_{2}=0.999$, and an initial learning rate of 0.0001.
Batch size is set to 2.
The scale factors $\lambda$ in Equation \ref{eq:our_loss} is set to 1.0, \textit{i.e.}, we do not tune $\lambda$.
For the meta-smoothing module, each filter of $\bm{W}^{r}_{u}$ and $\bm{W}_{s}$ is set to have the same dimension.
Specifically, we set $H_{u}=H_{s}=W_{u}=W_{s}=5$.

\subsection{Dataset}
In this work, we adopt two datasets for training and testing the proposed DNN-based stylizer, respectively. We train the stylization model on MS-COCO \cite{lin2014microsoft}, which contains 328k images covering 91 different object types. Both  content images and style images are sampled from MS-COCO for training. We also download 100 videos of different scenes from Videvo.net \cite{videonet} for evaluation, which contains about 41,500 frames. One style image is assigned to each video for performing style transformation.

\subsection{Stylization Speed}
The efficiency issue is one major concern about photorealistic style transfer. Therefore, we evaluate the efficiency of our proposed style transfer method by comparing with the methods proposed by Luan \textit{et al.} \cite{Luan_CVPR16} and Li \textit{et al.} \cite{Li_ECCV18}. Table \ref{table:exp_run_time_comp} shows the average time to perform style transfer on one frame with different resolutions. The numbers reported in Table \ref{table:exp_run_time_comp} are obtained by averaging the stylization time of 1000 frames which are randomly sampled from the testing data. Overall, our proposed method outperforms the methods proposed by Luan \textit{et al.}  and Li \textit{et al.}  with any resolution setting, and we can obtain greater speedup with increasing resolution. For example, the time of processing a $512\times256$ frame by our method is 356.7 times and 5.6 times faster than the approaches of Luan \textit{et al.}  and Li \textit{et al.}, respectively. Besides, for performing style transfer on a $1920\times1080$ frame, the two compared methods need over 1000 seconds and 38.72 seconds separately, but our method only costs 1.51 seconds, which is greatly faster.
In summary, benefiting from our proposed meta-smoothing module of the stylizer, our proposed model can perform stylization in a much more efficient way than existing work.

\begin{table}[ht]
\centering
\caption{Average run time (in seconds) comparison between existing photorealistic style transfer methods and ours (on an NVIDIA TITAN X Pascal).}
\label{table:exp_run_time_comp}
\begin{tabular}{rcccc}
\toprule
\textbf{Method}      & 512x256 & 768x384 & 1024x512 & 1920x1080\\
\midrule
Luan \textit{et al.}~\cite{Luan_CVPR16}  & 186.52  & 380.82  & 650.45 & $>$1000.00\\
Li \textit{et al.}~\cite{Li_ECCV18}      & 2.95    & 7.05    & 13.16 & 38.72\\

\textbf{Ours}  & \textbf{0.52}  & \textbf{0.73} & \textbf{0.99} & \textbf{1.51}\\
\bottomrule
\end{tabular}
\end{table}

\subsection{Speedup by Optical-flow-based Frame Interpolation}
As described in Section \ref{sec:design}, we design an optical-flow-based interpolation algorithm to interpolate stylized intermediate frames for improving efficiency. Therefore, we evaluate the speedup by comparing the run time of performing stylization by our pre-trained DNN model on an edge server with that of interpolating stylized intermediate frames on the mobile phone. Table \ref{tab:speedup} shows average run time of processing 1000 key frames and 1000 intermediate frames by those two methods separately. It demonstrates that the optical-flow-based interpolation method significantly outperforms performing DNN-based stylization in terms of efficiency, even though the run time of both methods will be increased with higher resolution frame.
In particular, performing DNN-based stylization on a $512\times256$ frame needs 0.52 seconds, but it only takes 0.00006 seconds for interpolating a stylized intermediate frame with the same resolution, which achieves about 866.7 times speedup. More important, since mobile phones usually record a video in high quality today, we also make the evaluation for high-resolution videos. Specifically, for processing a $1920\times1080$ frame, the optical-flow-based interpolation method only costs 0.02 seconds but the DNN-based stylization requires 1.51 seconds, indicating 75.5 times speedup. It can be imagined how slow it will be to perform the DNN-based stylization frame by frame. For instance, given a 10-minute video with resolution of $1920\times1080$ and frame rate at 30 fps, it will cost 7.55 hours to perform DNN-based stylization frame by frame even on the edge server.
\begin{table}[ht]
    \centering
    \caption{Average run time (in seconds) comparison between DNN-based stylization and optical-flow-based interpolation}
    \begin{tabularx}{0.47\textwidth}{cccc}
    \toprule
   
     \parbox{1.2cm}{\centering\textbf{Resolution}} & \parbox{2cm}{\centering\textbf{Stylization per frame (edge server)}} & \parbox{2cm}{\centering\textbf{Interpolation per frame (mobile)}} & \parbox{1.2cm}{\centering\textbf{Speedup}} \\
     \midrule
       512x256  & 0.52 & 0.0006 & 866.7\\
       768x384 & 0.73 & 0.002 & 365 \\
       1024x512 & 0.99 & 0.006 & 165 \\
       1920x1080 & 1.51 & 0.02 &75.5\\
       \bottomrule
    \end{tabularx}
    
    \label{tab:speedup}
\end{table}

\begin{figure}[ht]
    \centering
    \includegraphics[scale=0.4]{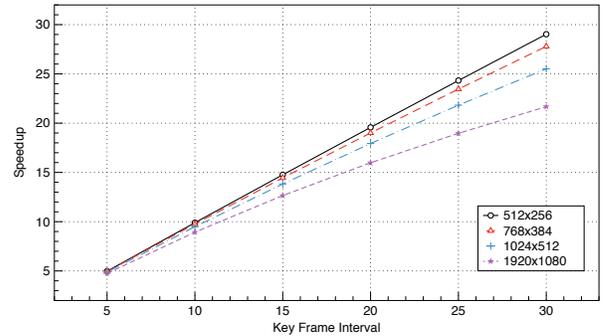}
    \caption{The speedup with different key frame intervals}
    \label{fig:speedup}
\end{figure}

In addition to evaluate the speedup for a single frame, we also quantitatively evaluate the speedup for performing stylization on videos. Benefiting from the optical-flow-based interpolation, we can define the speedup as Equation \ref{eq:speedup}:

\begin{equation}
\footnotesize
\begin{split}
    \text{speedup}&=\frac{\#\text{all frames}\times t_d}{\#\text{key frames}\times t_d+\#\text{intermediate frames}\times t_i}\\\
    &\approx \frac{\#\text{all frames}\times t_d}{\#\text{key frames}\times t_d}=\frac{\#\text{all frames}}{\#\text{key frames}}\\
    &\propto\text{key frame interval}\label{eq:speedup}
\end{split} 
\end{equation}

where $t_d$ represents the  time of performing stylization on a frame by the DNN-based stylizer, and $t_i$ expresses the  time of stylizing a frame with optical-flow-based interpolation algorithm, which are shown in Table \ref{tab:speedup}. Since $t_d\gg t_i$, speedup is approximately proportional to  the \textit{key frame interval} as shown Equation \ref{eq:speedup}. The key frame interval  is defined as how often a key frame appears in a particular video as Equation \ref{tab:speedup}.  In this experiment, we consider a video clip with 300 frames with different resolutions as an example, and the speedup with various key frame interval are displayed in Figure \ref{fig:speedup}. Generally, the higher key frame interval, the greater speedup due to less key frames in the video. Regarding with different resolutions, given the same key frame interval, since the higher resolution cost more time to perform style transfer for both key frames and intermediate frames, the speedup has a slight decrease but is  still directly proportional  to corresponding key frame interval.

\subsection{Quantitative Evaluation of Stylization Results}
In addition to the efficiency evaluation, we also experimentally evaluate the visual result of style transfer, including measuring the quality of stylization by the DNN-based stylizer and comparing the similarity between frames that are stylized by DNN-based stylizer with those interpolated by the optical-flow-based interpolation algorithm.

Firstly, we randomly sample 1000 frames with the resolution $512\times256$ from the testing data, and we compare the visual quality of stylized results based on those 1000 frames which are transformed by our DNN-based stylizer with those processed by the state-of-the-art photorealistic style transfer method ~\cite{Li_ECCV18}. In this experiment, we adopt two widely applied quantitative evaluation metrics \textit{Inception score} \cite{Salimans_NIPS16_2} and \textit{Fr\'echet Inception Distance (FID)} \cite{Salimans_NIPS16_2} for our evaluation, which are designed to measure two aspects of synthesized image quality: photorealism and diversity. Note that the bigger Inception score indicates higher visual quality while the smaller FID representing better image quality. The averaged result are reported in Table \ref{tab:quality}, demonstrating our method can generate visually better stylized results than the state-of-the-art.

\begin{table}[ht]
\centering
\caption{Visual quality comparison of stylized frames between the state-of-the-art~\cite{Li_ECCV18} and ours}
\label{table:exp_fid_comp}

\begin{tabular}{rcc}
\toprule
\textbf{Method}  & \textbf{Inception score} & \textbf{FID}\\
\midrule
Li \textit{et al.}~\cite{Li_ECCV18}      & 129.54 & 168.71 \\
\textbf{Ours}   & \textbf{135.20} & \textbf{164.23} \\
\bottomrule
\end{tabular}\label{tab:quality}
\end{table}

\begin{figure*}[ht]
    \centering
    \subfigure{\includegraphics[height=3cm,width=4cm]{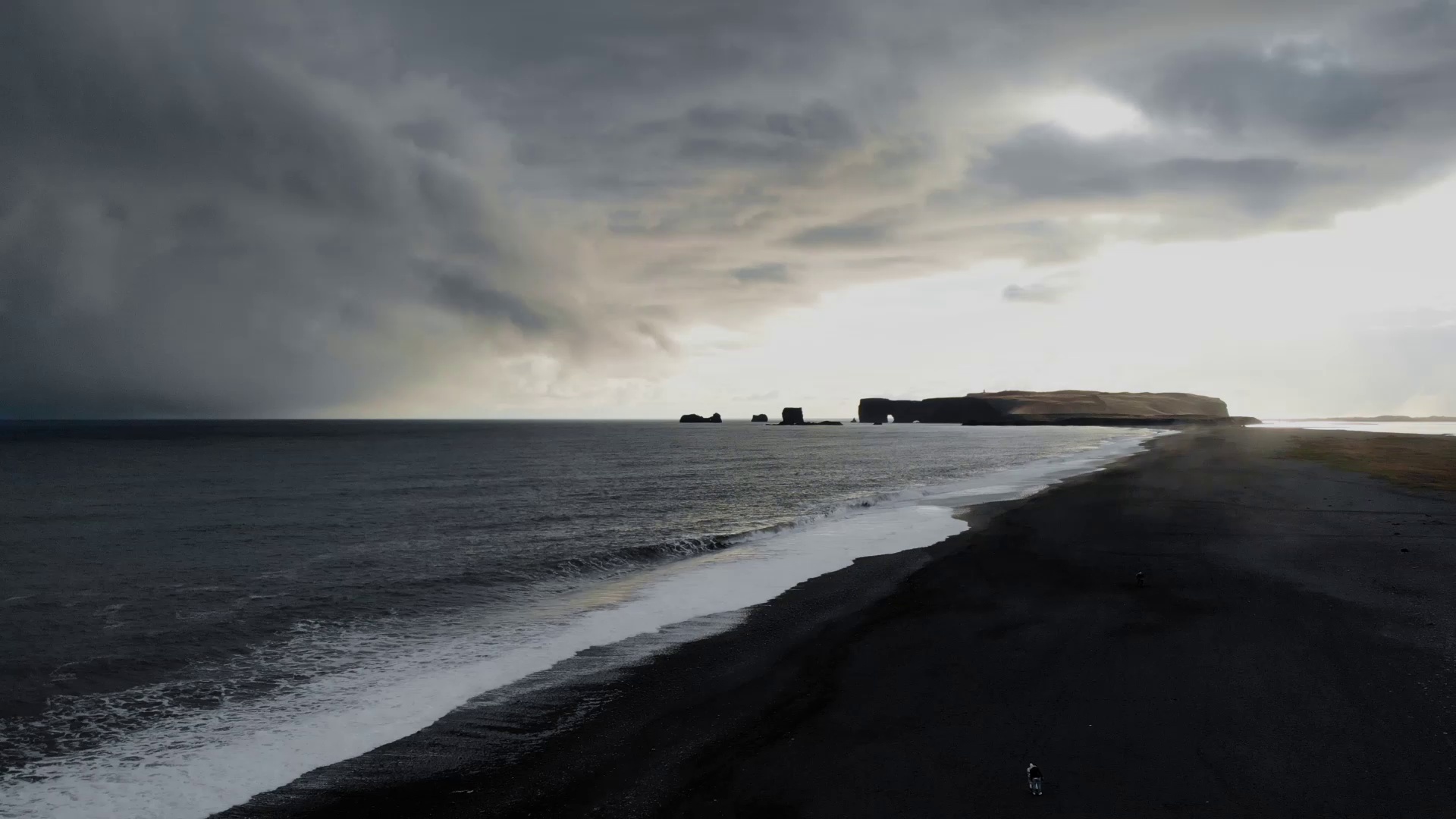}}\addtocounter{subfigure}{-1}
    \subfigure{\includegraphics[height=3cm,width=4cm]{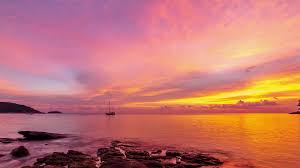}}\addtocounter{subfigure}{-1}
    \subfigure{\includegraphics[height=3cm,width=4cm]{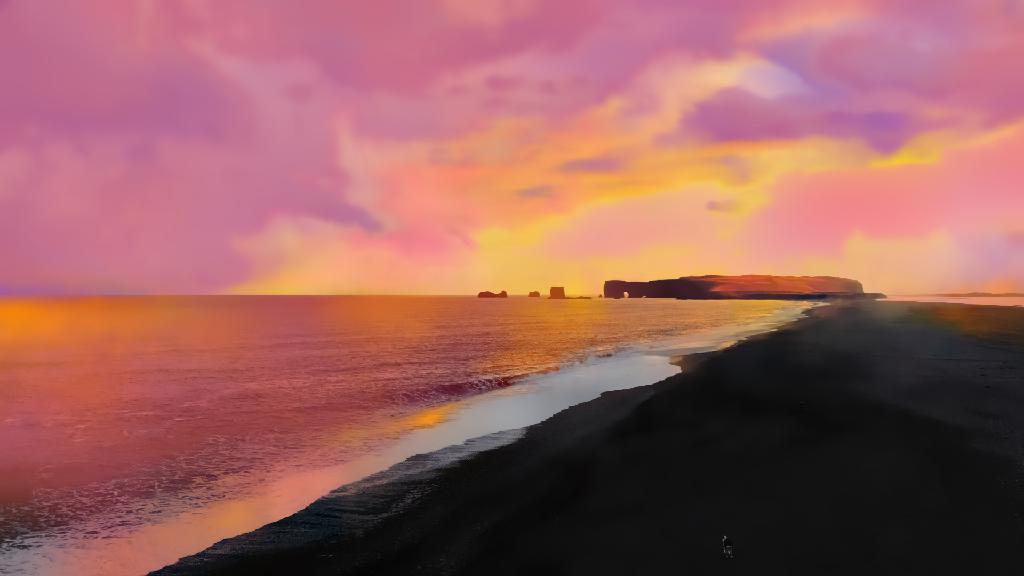}}\addtocounter{subfigure}{-1}
    \subfigure{\includegraphics[height=3cm,width=4cm]{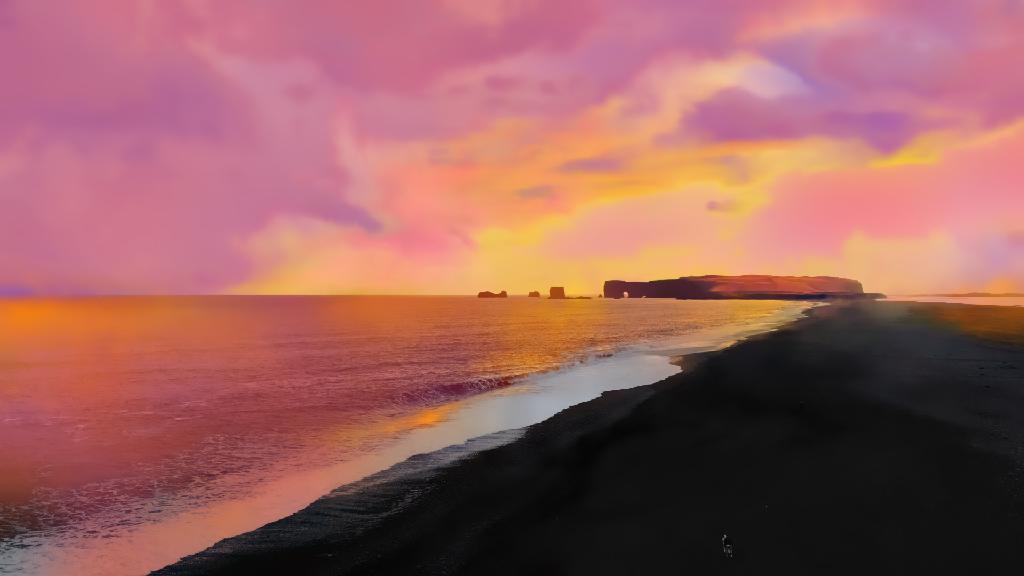}}\addtocounter{subfigure}{-1}
    \subfigure{\includegraphics[height=3cm,width=4cm]{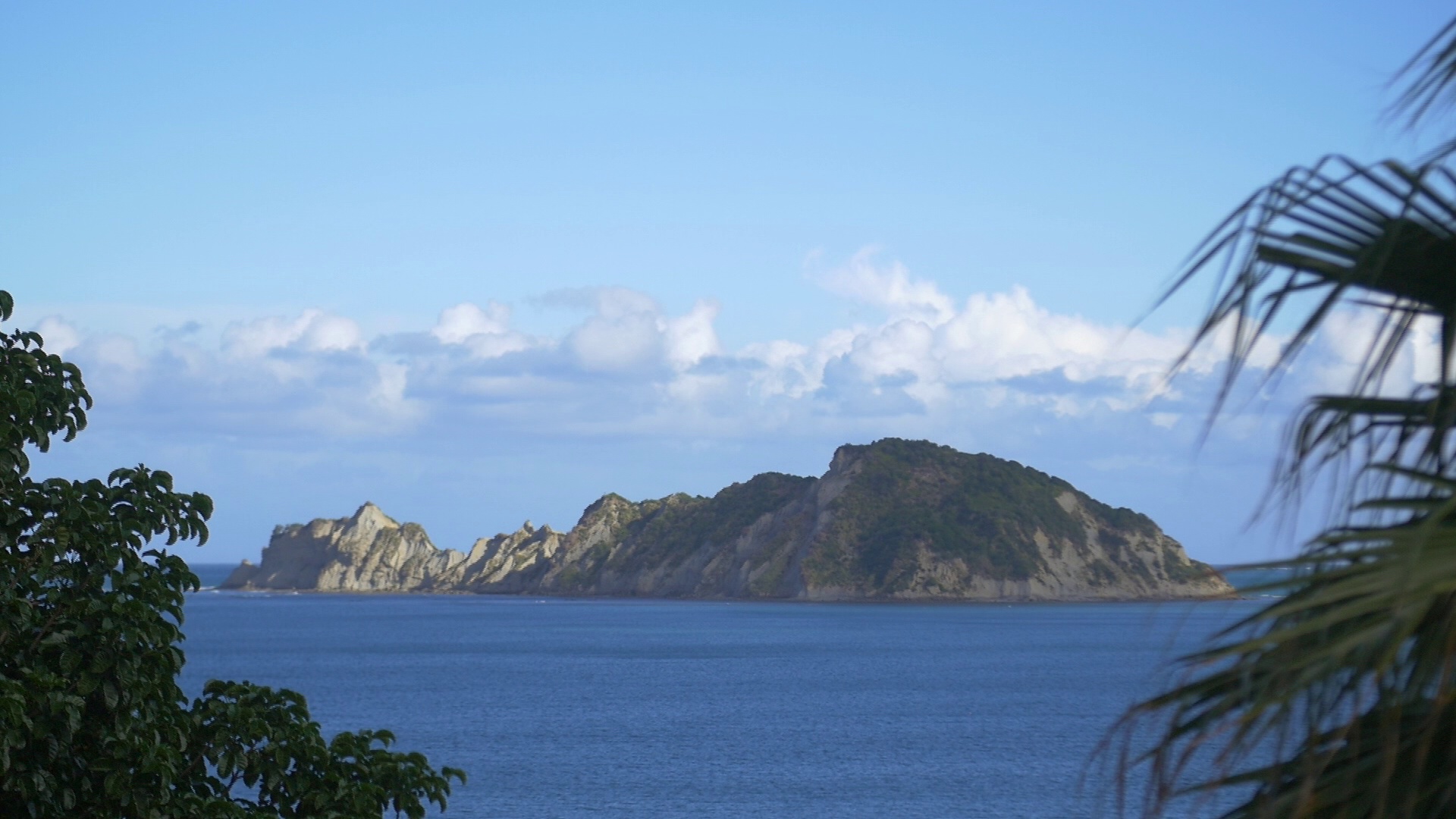}}\addtocounter{subfigure}{-1}
    \subfigure{\includegraphics[height=3cm,width=4cm]{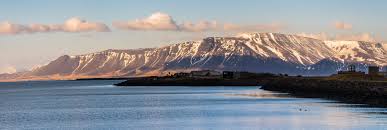}}\addtocounter{subfigure}{-1}
    \subfigure{\includegraphics[height=3cm,width=4cm]{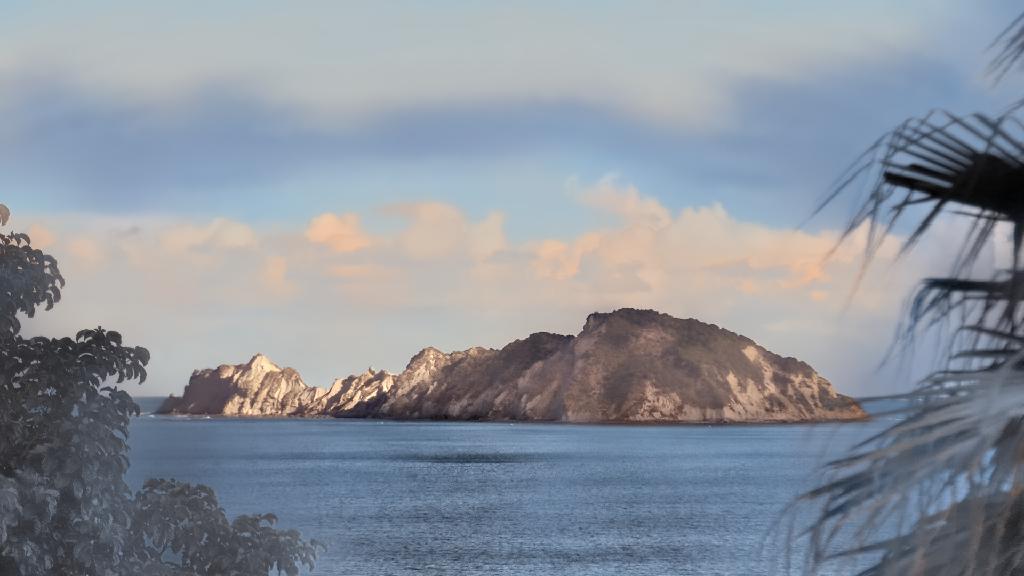}}\addtocounter{subfigure}{-1}
    \subfigure{\includegraphics[height=3cm,width=4cm]{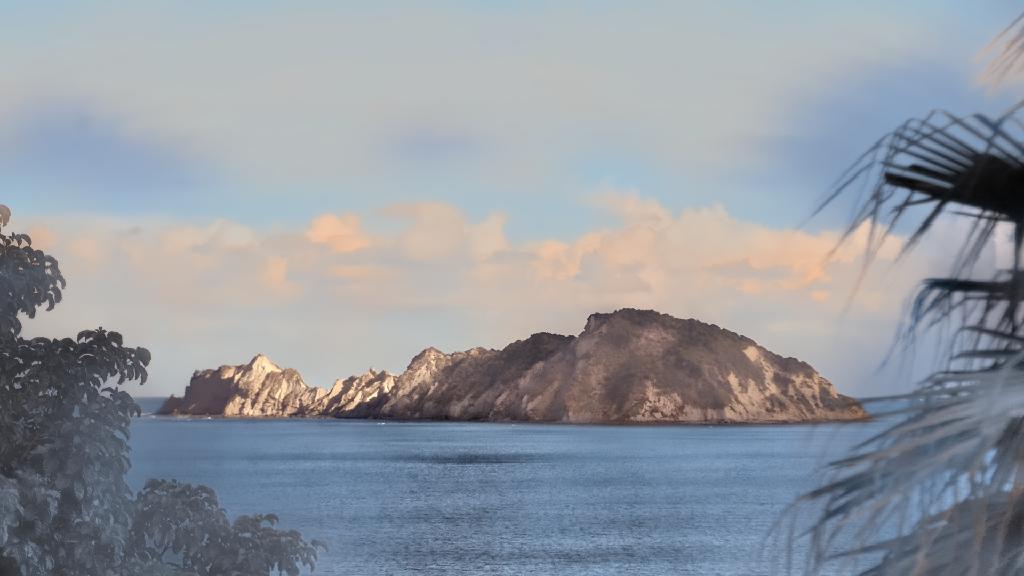}}
    \addtocounter{subfigure}{-1}
    \subfigure[input key frame]{\includegraphics[height=3cm,width=4cm]{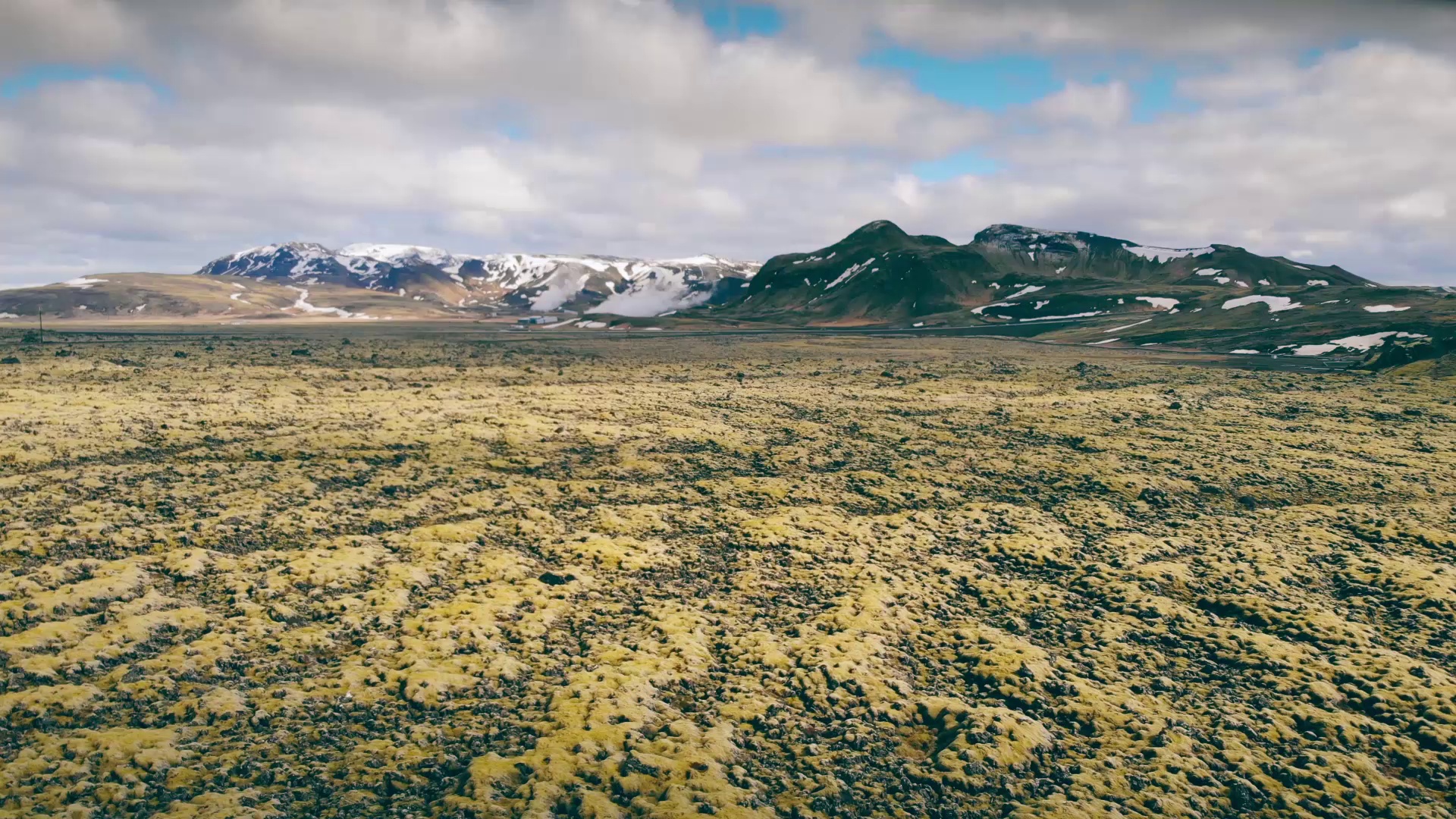}}
    \subfigure[style image]{\includegraphics[height=3cm,width=4cm]{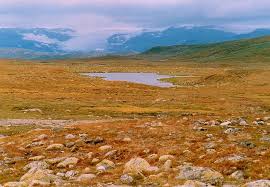}}
    \subfigure[stylized key frame]{\includegraphics[height=3cm,width=4cm]{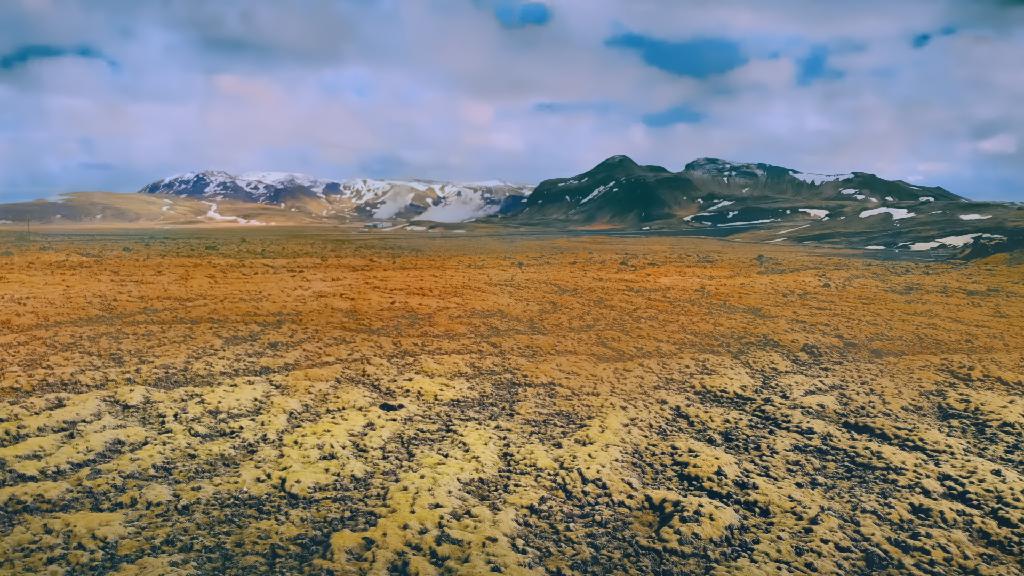}}
   \subfigure[interpolated intermediate frame]{\includegraphics[height=3cm,width=4cm]{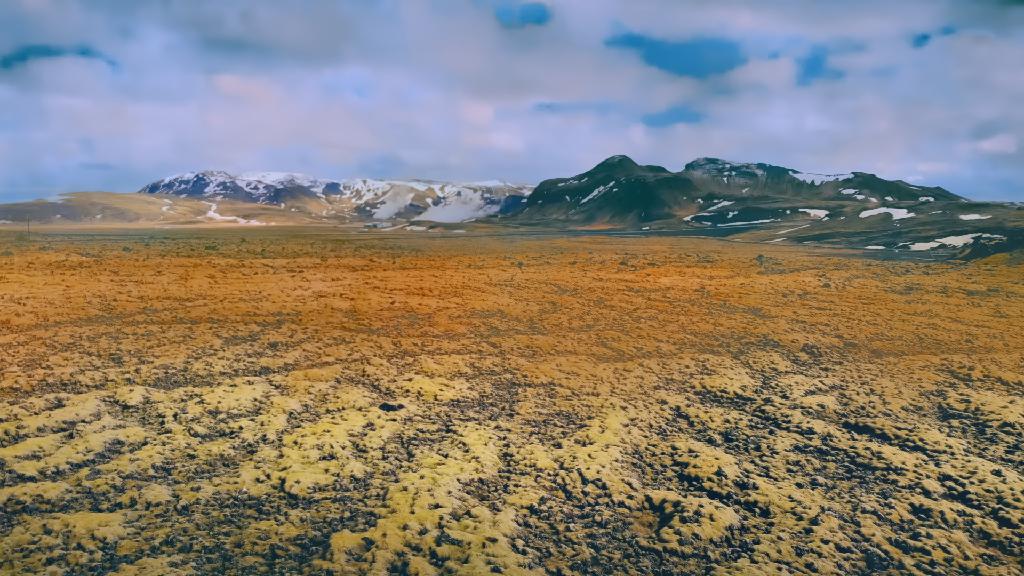}}
    \caption{Examples of stylized key frames and interpolated intermediate frames}
    \label{fig:stylization}
\end{figure*}

Besides, we also evaluate the visual quality of interpolated stylized intermediate frames. Ideally, we expect the visual quality of interpolated frames can be as close as frames which are directly processed by the DNN-based stylizer. Figure \ref{fig:stylization} shows several examples of stylized key frames by DNN-based stylizer and interpolated intermediate frames. In this example, we choose the intermediate frame that is next to the corresponding key frame for demonstration. As Figure \ref{fig:stylization} illustrates, both the stylized key frames and interpolated intermediate frames are successfully rendered into the target style while maintaining original content structure, but it is difficult to perceptually tell the difference between them in terms of visual quality. Furthermore, we also quantitatively evaluate image similarity between the stylized key frames and interpolated intermediate frames  by attempting to predict human perceptual similarity judgments. In this experiment, we adopt the widely applied metric \textit{multi-scale structural similarity (MS-SSIM)} \cite{wang2003multiscale,ma2016group} for measuring the frame similarity. MS-SSIM is a multi-scale perceptual similarity metric that attempts to pay less attention to aspects of an image that are not important for human perception. MS-SSIM values range between 0 and 1. The higher MS-SSIM values, the more perceptually similar between compared images. Specifically, we randomly choose 1000 frames from the testing data, and a pair of stylized frames for each of those 1000 frames are generated by the DNN-based stylizer and optical-flow-based interpolation algorithm, respectively. In addition, we evaluate the MS-SSIM for those 1000 pairs with different resolutions settings which are the same as used in above experiments. Table \ref{tab:ssim} shows the averaged results for those 1000 pairs, MS-SSIM is greater than 0.98 for all resolution settings, indicating that the stylized frames generated by the optical-flow-based interpolation algorithm have the perceptually comparable visual quality with frames that are directly processed by the DNN-based stylizer. 

In summary, above experiments demonstrate MVStylizer can efficiently perform style transfer of videos while achieving even better visual quality compared to the state-of-the-art method.

\begin{table}[ht]
    \centering
    \caption{MS-SSIM of stylized frames processed by DNN-based stylization and optical-flow-based interpolation}
    \begin{tabular}{cc}
    \toprule
   
     \textbf{Resolution} & \textbf{MS-SSIM} \\
     \midrule
       512x256  & 0.9845 \\
       768x384 & 0.9841 \\
       1024x512 & 0.9847 \\
       1920x1080 & 0.9849 \\
       \bottomrule
    \end{tabular}
    
    \label{tab:ssim}
\end{table}

\subsection{Latency Comparison}
We quantitatively compare the latency of sending a key frame to an edge server with that of sending a key frame to a conventional cloud server. In this experiment, we test 1000 frames that are sent by a user from Boston. The edge server is located in New York, but the other two conventional cloud servers are located in Los Angles and Hong Kong, respectively. The average results are shown in Table \ref{tab:latency}. Generally, the latency of send a key frame the edge server is about 10 times and 30 times lower than that of sending the key frame to the cloud server in Los Angles and in Hong Kong, respectively.
For example, it takes 0.031 seconds to upload a 1920$\times$1080 key frame to the edge server from the mobile user, but requires 0.318 seconds and 0.925 seconds to send the key frame to the cloud server in Los Angles and Hong Kong, respectively. The above results demonstrate that the edge server is able to provide the style transfer service to the mobile user with a significantly lower latency compared to the conventional cloud server.

\begin{table}[ht]
    \centering
    \caption{Average latency (in seconds) comparison between sending one frame to the edge server and to the cloud server.}
    \begin{tabularx}{0.47\textwidth}{cccc}
    \toprule
   
     \parbox{1cm}{\centering\textbf{Resolution}} & \parbox{2cm}{\centering\textbf{Edge Server (New York)}} & \parbox{2cm}{\centering\textbf{Cloud Server (Los Angles)}} & \parbox{2cm}{\centering\textbf{Cloud Server (Hong Kong)}} \\
     \midrule
       512x256  & 0.003 & 0.028 & 0.088\\
       768x384 & 0.006 & 0.058 & 0.176 \\
       1024x512 & 0.011 & 0.105 & 0.312 \\
       1920x1080 & 0.031 & 0.318 & 0.925\\
       \bottomrule
    \end{tabularx}
    
    \label{tab:latency}
\end{table}

\subsection{Performance Improvement by Edge-Cloud Federated Learning}\label{subsec:eval_federated}
\begin{figure*}
    \centering
    \subfigure[total loss]{\includegraphics[scale=0.33]{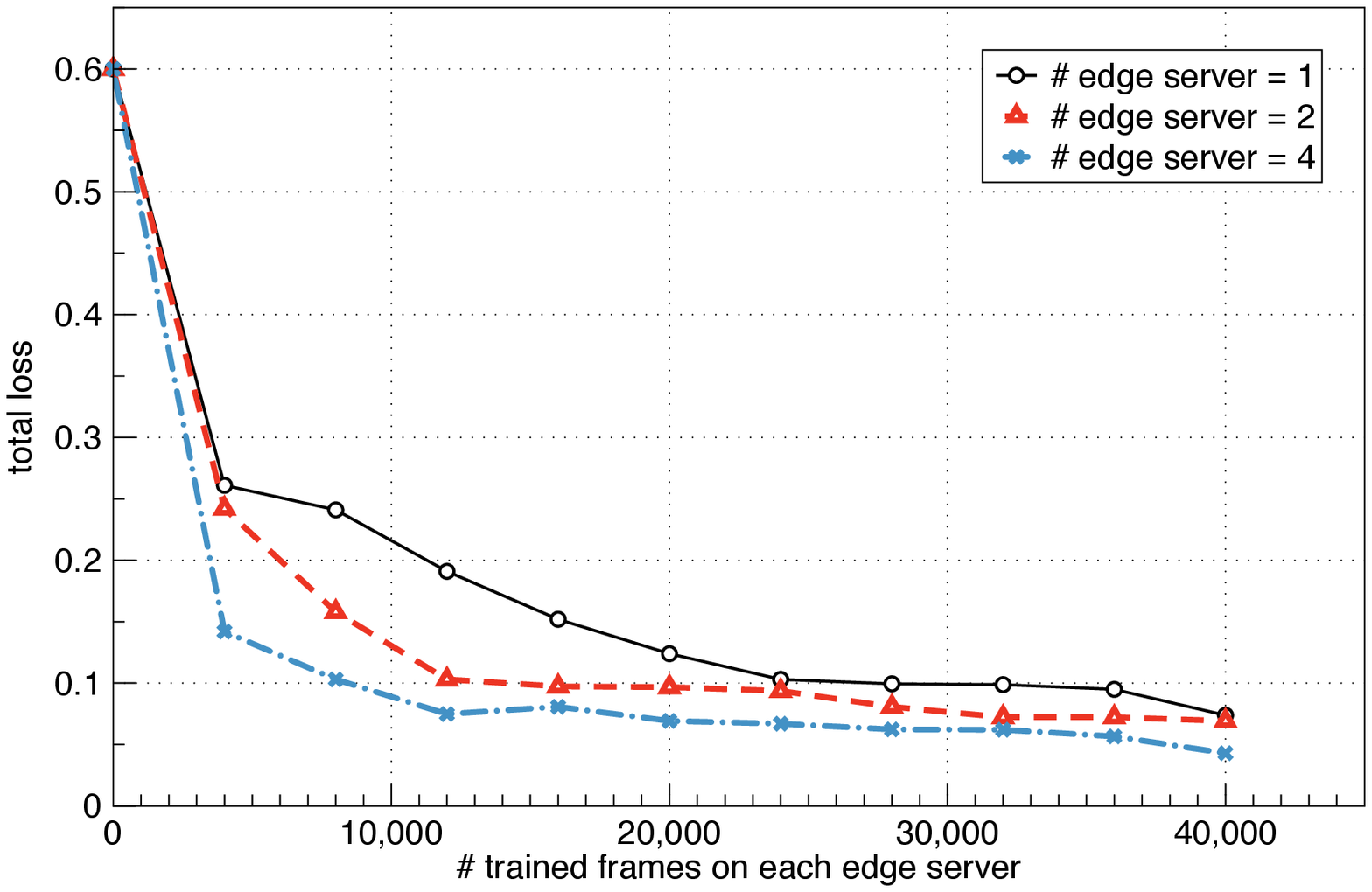}}
    \subfigure[content loss]{\includegraphics[scale=0.33]{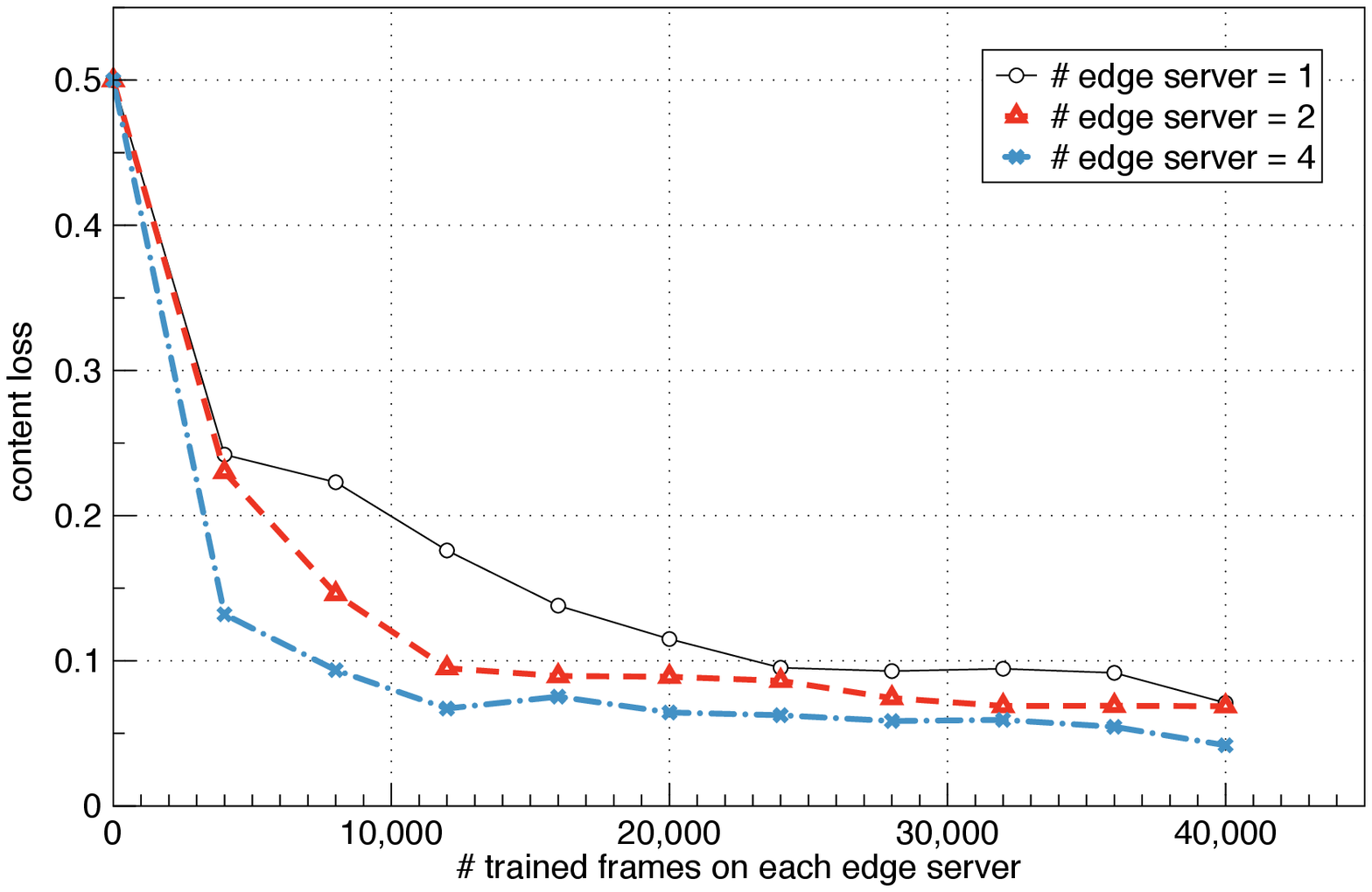}}
    \subfigure[style loss]{\includegraphics[scale=0.33]{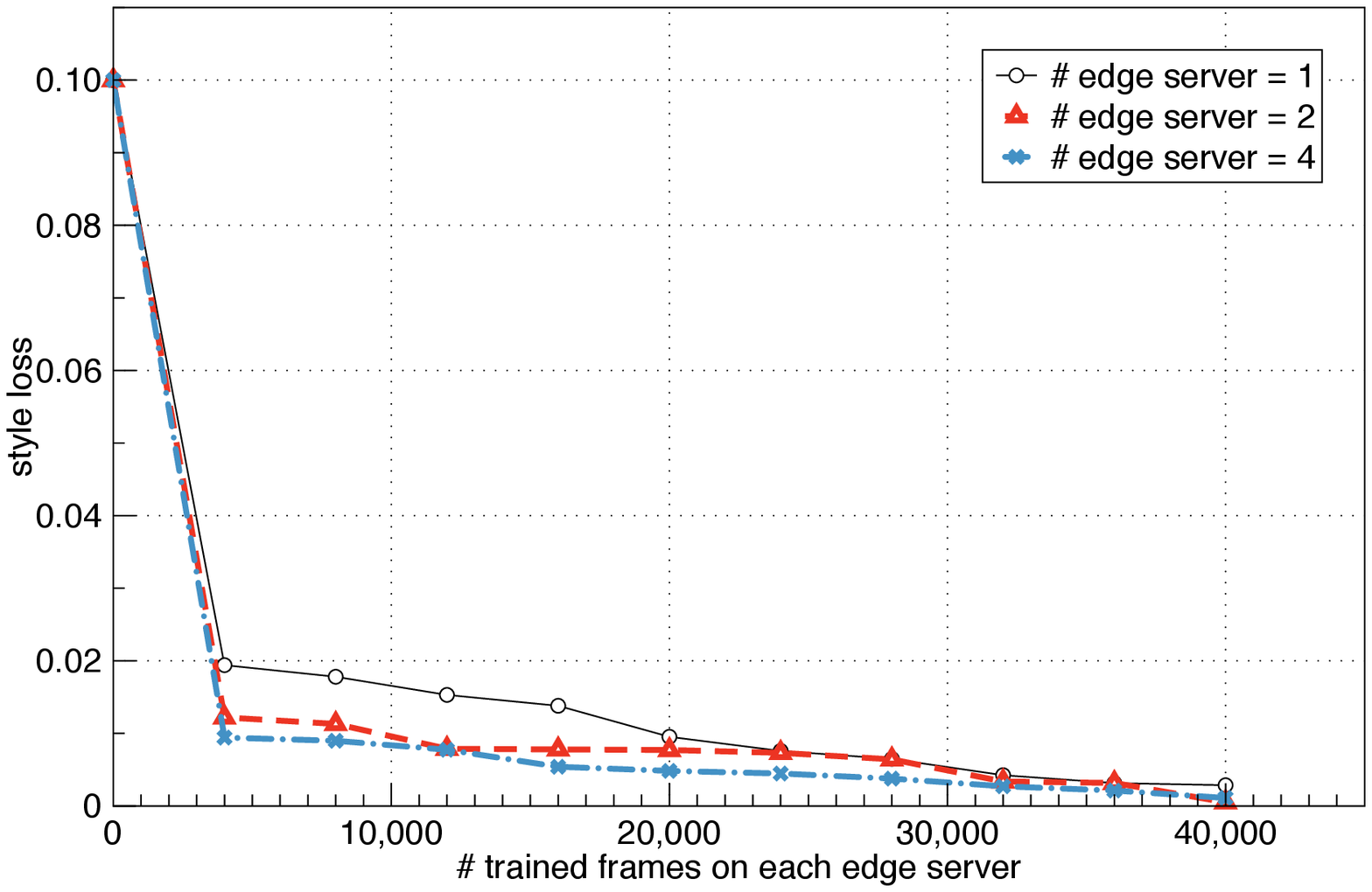}}
    \caption{Edge-cloud federated learning curves of total loss, content loss and style loss (\# participated edge server =1,2,4 )}
    \label{fig:loss}
\end{figure*}

As described in Section \ref{sec:design}, we apply an edge-cloud federated learning scheme to continuously improve the DNN-based stylizer on each edge server. In this experiment, we  quantitatively evaluate how edge-cloud federated learning scheme can improve the performance of DNN-based stylizer on each edge server through simulations. To make simulations, we enforce each participated edge server to sync the edge DNN with the cloud DNN after training on 4000 images that are randomly sampled from the training data, and the cloud DNN will make an update once receive all the synchronized parameters from all participated edge server. Then, the latest cloud DNN will be distributed to each edge server as the new edge DNN. In addition, we also change the number of participated edge servers to explore how it will affect the performance improvement. We evaluate the performance of the model based on the loss function defined as Equation \ref{eq:our_loss}, including the content loss and style loss. Figure \ref{fig:loss} shows the federated learning curves during the continuously training with different number of participated edge servers. In general, the more participated edge servers, the faster and greater the performance can be improved. For example, if there are 4 participated edge servers, the total loss can be reduced to 0.0748 after 12000 images are trained on each edge server. However, it requires to train 40000 images when there is only one participated edge server for achieving the same performance, and 32000 images on each of the two participated edge servers. The results show the edge-cloud federated learning can more efficiently improve the performance with an increasing number of participated edge servers.

\section{Related Work}\label{sec:related}


\textbf{Neural Style Transfer.}
Current neural style transfer techniques fit one of two mainstreams \cite{2017arXiv170504058J}, \textit{image-optimization-based online neural methods} and \textit{model-optimization-based offline neural methods}. Generally, the first category transfers the style by optimizing an image in an iterative way, while the second category aims to optimize a generative model offline which can generate the stylized image with a single forward pass. Gatys \textit{et al.} \cite{gatys2015neural,gatys2016image} proposed a seminal work demonstrating the power in style transfer by separating and recombining image content and style. It is inspired by observing that a CNN can separately extract content information from an original image and style information from a style image. Based on such observations, a CNN model can be trained to recombine the extracted content and style information to generate the target stylized image. However, this method only compares the content and stylized image in the feature space, which inevitably lose some low-level information contained in the image that can lead to distortion and abnormal artistic effects of stylized outputs. To preserve the structure coherence, Li \textit{et al.} \cite{li2017laplacian} introduced an additional Laplacian loss to constrain  low-level features in pixel space. Although image-optimization-based can achieve impressive stylized results, they have a common limitation in efficiency. To address the efficiency issue, various model-optimization-based offline neural methods have been proposed. Johnson \textit{et al.} \cite{johnson2016perceptual} and Ulyanov \textit{et al.} \cite{ulyanov2016texture} proposed the first two model-optimization-based algorithms for style transfer, which generate stylized result with a single forward pass through a pre-trained style-specific CNN. Even though those algorithms can achieve real-time style transfer, they require separate generative networks to be trained for each specific style, which is quite inflexible. Therefore, multiple-style-per-model neural methods are proposed to improve the flexibility by integrating multiple styles into one single model by tuning a small number of parameters for each style \cite{Dumoulin:2017te,Chen:2017tx}  or combining both style and content as inputs to the generative model \cite{2017arXiv170301664L,:2018wv}. Furthermore, several works have been done for designing one single mode transfer arbitrary artistic styles by exploiting texture modeling techniques \cite{Chen:wc,Ghiasi:2017wz}.

\textbf{Video Style Transfer.} Compared with above image style transfer techniques, video style transfer algorithms need to consider the smooth transition between consecutive frames. Ruder \textit{et al.} \cite{ruder2016artistic,Ruder:2018vx} introduced a temporal consistency loss based on optical flow for video style transfer. Huang \textit{et al.} \cite{huang2017real} designed an augmented temporal consistency loss by computing the outputs of style transfer network for two consecutive frames. A flow subnetwork was proposed by Chen \textit{et al.} \cite{chen2017coherent} to produce feature flow, which can be used to wraps feature activations from a pre-trained stylization encoder.

\textbf{Photorealistic Style Transfer.} Most existing works focus on artistic style transfer which can tolerate some distortion, but photorealistic style transfer requires more strict structure preservation of the content image. Luan \textit{et al.} \cite{luan2017deep} firstly proposed a two-stage optimization for photorealistic style transfer, which firstly renders a given photo with non-photorealistic style and then penalizes image distortions by adding a photorealistic regularization. However, this algorithm is very computational expensive. Mechrez \textit{et al.} \cite{Mechrez:uc} also adopts above two-stage optimization scheme, but they refine the photorealistic rendering effect by matching the gradients in the stylized image to those in the content image. To improve the efficiency issue, Li \textit{et al.} \cite{Li_ECCV18} designed a two-step photorealistic style transfer algorithm, including the \textit{stylization step} and \textit{smoothing step}. The stylization step aims to generate stylized output based on existing neural style transfer algorithms but replaces upsampling layers with unpooling layer for less distortion, and then the smoothing step is applied to remove structural artifacts.

Even though there exist some works on video style transfer and photorealistic style transfer, none of them is specifically designed for performing photorealistic style transfer of videos on resource-constrained devices, such as mobile phones.
\section{Conclusion} \label{sec:conclusion}
In this paper, we designed MVStylizer to efficiently perform photorealistic style transfer of videos on mobile phones with the assistance of an edge server. Considering the stylization is very computational expensive, we proposed an optical-flow-based interpolation algorithm, so that only key frames in the video need to be uploaded to the edge server where they can be processed by the pre-trained DNN-based stylizer and the rest of stylized intermediate frames can be interpolated based on the pre-computed optical flow information from the original video and stylized key frames. A meta-smoothing module is also designed in the DNN-based stylizer for improving the efficiency of performing style transfer on the edge server. In addition, we adopt an edge-cloud federated learning scheme to continuously enhancing the performance of DNN-based stylizer.  Experiments demonstrate 75.5 times speedup compared with performing style transfer frame by frame using the DNN-based stylizer even with the high resolution, while generating the stylized videos with even better visual quality compared to the state-of-the-art method. Furthermore, it also demonstrates the edge-cloud federated learning scheme can facilitate in continuously improving the performance of the DNN-based stylizer in an efficiency way.

\bibliographystyle{ACM-Reference-Format}
\bibliography{refs}
\end{document}